\pdfoutput=1
\documentclass[useAMS,usenatbib]{mn2e}
\usepackage{apjfonts}
\usepackage{graphicx}
\usepackage{amsmath}
\usepackage{amssymb}
\usepackage{ctable}
\usepackage{fixltx2e}
\usepackage{hyperref}
\hypersetup{colorlinks=true,linkcolor=blue,citecolor=blue,filecolor=blue,urlcolor=blue}

\newcommand{\be}{\begin{equation}}
\newcommand{\ee}{\end{equation}}
\newcommand{\ba}{\begin{eqnarray}}
\newcommand{\ea}{\end{eqnarray}}

\newcommand{\Ms}{M_{\ast}}
\newcommand{\Msun}{{\rm M}_{\sun}}
\newcommand{\Zsun}{Z_{\sun}}
\newcommand{\Sg}{\Sigma_g}
\newcommand{\Ss}{\Sigma_{\ast}}
\newcommand{\DSs}{\dot{\Sigma}_{\ast}}

\newcommand{\nc}{n_{\rm th}}
\newcommand{\cm}{{\rm cm}}
\newcommand{\km}{{\rm km}}
\newcommand{\pc}{{\rm pc}}
\newcommand{\kpc}{{\rm kpc}}
\newcommand{\dex}{{\rm dex}}
\newcommand{\dkpc}{\dex\,\kpc^{-1}}
\newcommand{\s}{{\rm s}}
\newcommand{\yr}{{\rm yr}}

\newcommand{\Gyr}{{\rm Gyr}}
\newcommand{\dd}{{\rm d}}

\newcommand{\hop}{\citet{hopkins.2014:fire.galaxy}}
\newcommand{\cafg}{\citet{faucher.2015:fire.neutral.hydrogen}}
\newcommand{\chan}{\citet{chan.2015:fire.cusp.core}}
\newcommand{\feld}{\citet{feldmann.2016:fire.quenching.letter}}
\newcommand{\hafen}{\citet{hafen.2016:fire.lyman.limit}}

\newcommand{\aj}{AJ}
\newcommand{\apj}{ApJ}
\newcommand{\apjl}{ApJ}
\newcommand{\apjs}{ApJS}
\newcommand{\mnras}{MNRAS}
\newcommand{\aap}{A\&A}
\newcommand{\araa}{ARA\&A}
\newcommand{\nat}{Nature}

\title[Gas-phase Metallicity Gradients in High-redshift Galaxies]
{Why do high-redshift galaxies show diverse gas-phase metallicity gradients?}

\author[X. Ma et al.]{
  \parbox[t]{1.0\textwidth}{
   Xiangcheng Ma,$^1$\thanks{E-mail: \href{mailto:xchma@caltech.edu}{xchma@caltech.edu}}
   Philip F. Hopkins,$^1$
   Robert Feldmann,$^{2,3}$
   Paul Torrey,$^{1,4}$ \\
   Claude-Andr{\'e} Faucher-Gigu{\`e}re$^5$ and
   Du{\v s}an Kere{\v s}$^6$
  }
  \vspace{5pt} \\
  $^1$TAPIR, MC 350-17, California Institute of Technology, Pasadena, CA 91125, USA \\ 
  $^2$Institute for Computational Science, University of Zurich, Zurich CH-8057, Switzerland \\
  $^3$Department of Astronomy and Theoretical Astrophysics Center, University of California Berkeley, Berkeley, CA 94720 \\
  $^4$MIT Kavli Institute for Astrophysics \& Space Research, Cambridge, MA, 02139, USA \\
  $^5$Department of Physics and Astronomy and CIERA, Northwestern University, 2145 Sheridan Road, Evanston, IL 60208, USA \\
  $^6$Department of Physics, Center for Astrophysics and Space Sciences, University of California at San Diego, 9500 Gilman Drive, La Jolla, CA 92093
}

\pagerange{\pageref{firstpage}--\pageref{lastpage}}
\date{Draft version \today}

\begin{document}
\maketitle
\label{firstpage}

\begin{abstract}
Recent spatially resolved observations of galaxies at $z\sim0.6$--3 reveal that high-redshift galaxies show complex kinematics and a broad distribution of gas-phase metallicity gradients. To understand these results, we use a suite of high-resolution cosmological zoom-in simulations from the Feedback in Realistic Environments (FIRE) project, which include physically motivated models of the multi-phase ISM, star formation, and stellar feedback. Our simulations reproduce the observed diversity of kinematic properties and metallicity gradients, broadly consistent with observations at $z\sim0$--3. Strong negative metallicity gradients {\em only} appear in galaxies with a rotating disk, but not {\em all} rotationally supported galaxies have significant gradients. Strongly perturbed galaxies with little rotation always have flat gradients. The kinematic properties and metallicity gradient of a high-redshift galaxy can vary significantly on short time-scales, associated with starburst episodes. Feedback from a starburst can destroy the gas disk, drive strong outflows, and flatten a pre-existing negative metallicity gradient. The time variability of a single galaxy is statistically similar to the entire simulated sample, indicating that the observed metallicity gradients in high-redshift galaxies reflect the instantaneous state of the galaxy rather than the accretion and growth history on cosmological time-scales. We find weak dependence of metallicity gradient on stellar mass and specific star formation rate (sSFR). Low-mass galaxies and galaxies with high sSFR tend to have flat gradients, likely due to the fact that feedback is more efficient in these galaxies. We argue that it is important to resolve feedback on small scales in order to produce the diverse metallicity gradients observed.
\end{abstract}


\begin{table*}
\caption{Simulation details.}
\centering
\begin{center}
\begin{tabular}{cccccccc}
\hline
Name & $M_{\rm halo}$ ($z=0$) & $M_{\rm halo}$ ($z=2$) & $m_b$ & $\epsilon_b$ & $m_{\rm dm}$ & $\epsilon_{\rm dm}$ & Reference \\
 & ($\Msun$) & ($\Msun$) & ($\Msun$) & ($\pc$) & ($\Msun$) & ($\pc$) & \\ 
\hline
m11 & 1.4e11 & 3.8e10 & 7.1e3 & 7.0 & 3.5e4 & 70 & \hop \\
m12v & 6.3e11 & 2.0e11 & 3.9e4 & 10 & 2.0e5 & 140 & \hop  \\ 
m12q & 1.2e12 & 5.1e11 & 7.1e3 & 10 & 2.8e5 & 140 & \hop  \\ 
m12i & 1.1e12 & 2.7e11 & 5.0e4 & 14 & 2.8e5 & 140 & \hop  \\ 
m13 & 6.0e12 & 8.4e11 & 3.6e5 & 21 & 2.2e6 & 210 & \hop \\
m11h383 & 1.6e11 & 4.1e9 & 1.7e4 & 10 & 8.3e4 & 100 & \chan \\
m11.4a & 2.6e11 & 8.9e10 & 3.3e4 & 9 & 1.7e5 & 140 & \hafen \\
m11.9a & 8.4e11 & 1.3e11 & 3.4e4 & 9 & 1.7e5 & 140 & \hafen \\
MFz0\_A2 & 1.0e13 & 2.2e12 & 3.0e5 & 9 & 1.4e6 & 140 & \hafen \\
z2h350 & -- & 7.9e11 & 5.9e4 & 9 & 2.9e5 & 143 & \cafg  \\
z2h400 & -- & 7.9e11 & 5.9e4 & 9 & 2.9e5 & 143 & \cafg \\
z2h450 & -- & 8.7e11 & 5.9e4 & 9 & 2.9e5 & 143 & \cafg \\
z2h506 & -- & 1.2e12 & 5.9e4 & 9 & 2.9e5 & 143 & \cafg \\
z2h550 & -- & 1.9e11 & 5.9e4 & 9 & 2.9e5 & 143 & \cafg \\
z2h600 & -- & 6.7e11 & 5.9e4 & 9 & 2.9e5 & 143 & \cafg \\
z2h650 & -- & 4.0e11 & 5.9e4 & 9 & 2.9e5 & 143 & \cafg \\
z2h830 & -- & 5.4e11 & 5.9e4 & 9 & 2.9e5 & 143 & \cafg \\
A1:0 & -- & 2.3e12 & 3.3e4 & 10 & 1.7e5 & 143 & \feld \\
A2:0 & -- & 2.9e12 & 3.3e4 & 10 & 1.7e5 &143 & \feld \\
A3:0 & -- & 2.4e12 & 3.3e4 & 10 & 1.7e5 & 143 & \feld \\
A4:0 & -- & 2.8e12 & 3.3e4 & 10 & 1.7e5 & 143 & \feld \\
A5:0 & -- & 2.3e12 & 3.3e4 & 10 & 1.7e5 & 143 & \feld \\
A6:0 & -- & 2.6e12 & 3.3e4 & 10 & 1.7e5 & 143 & \feld \\
A7:0 & -- & 2.5e12 & 3.3e4 & 10 & 1.7e5 & 143 & \feld \\
A8:0 & -- & 3.5e12 & 3.3e4 & 10 & 1.7e5 & 143 & \feld \\
A9:0 & -- & 2.8e12 & 3.3e4 & 10 & 1.7e5 & 143 & \feld \\
A10:0 & -- & 3.2e12 & 3.3e4 & 10 & 1.7e5 & 143 & \feld \\
B1:0 & -- & 8.3e12 & 3.3e4 & 10 & 1.7e5 & 143 & \feld \\
B2:0 & -- & 9.0e12 & 3.3e4 & 10 & 1.7e5 & 143 & \feld \\
B3:0 & -- & 9.7e12 & 3.3e4 & 10 & 1.7e5 & 143 & \feld \\
B4:0 & -- & 8.5e12 & 3.3e4 & 10 & 1.7e5 & 143 & \feld \\
B5:0 & -- & 9.1e12 & 3.3e4 & 10 & 1.7e5 & 143 & \feld \\
\hline
\multicolumn{8}{p{0.7\textwidth}}{Parameters describing the initial conditions for our simulations (units are physical):} \\
\multicolumn{8}{p{0.7\textwidth}}{(1) Name: Simulation designation.} \\
\multicolumn{8}{p{0.7\textwidth}}{(2) $M_{\rm halo}$: Approximate mass of the main halo (most massive halo), at $z=0$ and $z=2$.} \\
\multicolumn{8}{p{0.7\textwidth}}{(3) $m_b$: Initial baryonic (gas and star) particle mass in the high-resolution region.} \\ 
\multicolumn{8}{p{0.7\textwidth}}{(4) $\epsilon_b$: Minimum baryonic Plummer-equivalent force softening (minimum SPH smoothing lengths are comparable or smaller). Force softening is adaptive (mass resolution is fixed).}\\
\multicolumn{8}{p{0.7\textwidth}}{(5) $m_{\rm dm}$: Dark matter particle mass in the high-resolution region.} \\ 
\multicolumn{8}{p{0.7\textwidth}}{(6) $\epsilon_{\rm dm}$: Minimum dark matter Plummer-equivalent force softening (fixed in physical units at all redshifts).} \\
\multicolumn{8}{p{0.7\textwidth}}{(7) Reference: Where the simulation is first presented.} \\
\multicolumn{8}{p{0.7\textwidth}}{Note: Detailed physical properties of these galaxies are presented in Appendix \ref{sec:append:gal}.} \\
\end{tabular}
\end{center}
\label{tbl:sim}
\end{table*}%

\section{Introduction}
\label{intro}
Metallicity is a fundamental property of galaxies. In the local Universe, galaxy stellar mass correlates tightly with both gas-phase metallicity \citep[e.g.][]{tremonti.2004:gas.mzr.sdss,lee.2006:gas.mzr.local.dwarf} and stellar metallicity \citep[e.g.][]{gallazzi.2005:stellar.mzr.sdss,kirby.2013:stellar.mzr.dwarf}, known as the galaxy mass--metallicity relation (MZR). The MZR also exists at higher redshifts up to $z\sim3$ \citep[e.g.][]{erb.2006:mzr.z2,maiolino.2008:gas.mzr.zgt3,mannucci.2009:gas.mzr.z3pt1,zahid.2011:deep2.mzr.z0pt8,yabe.2014:mzr.z1pt4.fmos,steidel.2014:nebular.mosfire,sanders.2015:gas.mzr.z2pt3}. The MZR evolves smoothly with redshift, with galaxies being more metal-enriched at lower redshift \citep[e.g.][]{zahid.2013:mzr.evolve}. The MZR results from the interplay between gas accretion and recycling, star formation, and feedback-driven outflows \citep[e.g.][]{edmunds.1990:gas.flow.metal,dave.2012:bathtub.model,lilly.2013:analytic.model.fmr,feldmann.2013:cosmic.sf,lu.2015:metal.constrain.outflow}, so it is widely used to constrain feedback models in cosmological simulations and semi-analytic models of galaxy formation \citep[e.g.][]{dave.2011:cosmo.sim.metal,torrey.2014:illutris.scaling,lu.2014:sam.compare.candel,ma.2016:fire.mzr}.

Historically, galaxy metallicity is usually measured in the central regions despite the presence of  metallicity gradients. Since \citet{searle.1971:metal.gradient}, it has been known that galaxies in the local Universe tend to have negative gas-phase metallicity gradients, which means that galaxies are more metal-enriched in the central region than at the outskirt \citep[e.g.][]{zaritsky.1994:metal.grad.spiral,vanzee.1998:metal.grad.spiral,sanchez.2012:pings.ifu.nearby,sanchez.2014:califa.metal.grad}. The slope of metallicity gradients of non-interacting galaxies, if normalized to some characteristic radius (e.g. the effective radius), does not depend strongly on galaxy properties, such as morphology, the existence of bars, magnitude, stellar mass, etc. (e.g. \citealt{zaritsky.1994:metal.grad.spiral,sanchez.2014:califa.metal.grad,ho.2015:metal.grad.coevolve}; however, see \citealt{vilacostas.1992:grad.property}). This can be understood by a simple model where gas and stellar disks co-evolve under virtually closed-box assumptions \citep{ho.2015:metal.grad.coevolve}. Interacting galaxies are under-abundant in their central regions \citep[e.g.][]{kewley.2006:low.metal.pair,peeples.2009:low.metal.purturb} and show evidence of shallower gas-phase metallicity gradients than isolated galaxies of similar masses \citep[e.g.][]{vilacostas.1992:grad.property,kewley.2010:metal.grad.pairs,rupke.2010:metal.grad.interaction}, owing to strong radial inflow of low-metallicity gas from the outskirts toward the galactic center \citep[e.g.][]{rupke.2010:merger.sim.metal.grad,torrey.2012:metal.interacting}.

It is only in the past few years that gas-phase metallicity gradients have been directly measured in galaxies beyond the local Universe. Early attempts include resolved studies of several strongly lensed galaxies at redshift $z\sim1.5$--2.4 \citep[e.g.][]{yuan.2011:metal.gradient.spiral,jones.2010:zgrad.z2.inside.out,jones.2013:lense.metal.grad}. Four out of five of these galaxies show well-ordered rotation and have steeper slopes (in ${\rm dex\,kpc^{-1}}$) in metallicity gradient than those of galaxies in the local Universe. In addition, \citet{maciel.2003:mw.metal.grad.evolve} measured the abundances of planetary nebulae in the Milky Way (MW) generated by stars spanning a broad age interval and suggested that the MW had steeper metallicity gradients back to $z\sim1.5$. These results support the so-called ``inside-out'' growth model of galaxy formation \citep[e.g.][]{bouwens.1997:inside.out.growth}. In this scenario, the central galactic bulge formed rapidly at early times, building a steep radial metallicity gradient at high redshift. The size of the disk gradually grows with time via gas infall. The metallicity gradient gradually weakens via star formation in the outer disk and radial gas mixing. Such a picture is also seen in some cosmological hydrodynamic simulations \citep[e.g.][]{pilkington.2012:disk.inside.out,gibson.2013:metal.grad.notes}, where the metallicity gradients are steepest at high redshift and gradually flatten at late times.

Recently, \citet{leeth.2016:lense.metal.grad} have studied 11 gravitationally lensed galaxies at redshift $z\sim1.4$--2.5 and found a broad distribution of kinematics and abundance patterns \citep[see also][]{jones.2015:glass.metal.grad,wang.2016:glass.metal.grad}. Most galaxies in their sample show no features of well-ordered rotation and tend to have flat gas-phase metallicity gradient, in contrast to earlier statements that high-redshift galaxies tend to have stronger metallicity gradients \citep{jones.2013:lense.metal.grad}. Moreover, large samples of non-lensed galaxies at redshift $z\sim0.6$--3 also show diverse metallicity gradients \citep[e.g.][]{cresci.2010:metal.gradient,queyrel.2012:massiv.metal.grad,swinbank.2012:hizels.metal.grad,stott.2014:metal.grad.z1.kmos,wuyts.2016:kmos3d.metal.grad}, with slope varying from negative to flat and positive. For example, \citet{wuyts.2016:kmos3d.metal.grad} have found that only 15 out of 180 galaxies that have spatially resolved measurements of abundances in a sample of galaxies at $z\sim0.6$--2.7 show statistically significant non-zero slope of metallicity gradients. These results complicate the simple `inside-out' growth picture.

Various studies have pointed out the necessity of strong feedback in order to avoid steep metallicity gradients in high-redshift galaxies in cosmological hydrodynamic simulations \citep[e.g.][]{pilkington.2012:disk.inside.out,gibson.2013:metal.grad.notes,angles.2014:galactic.outflow}. For example, \citet{gibson.2013:metal.grad.notes} compared two cosmological simulations run with different feedback models and showed that their `enhanced' feedback model produces constantly flat metallicity gradients at high redshift, whereas their `conservative' feedback model tends to follow the simple `inside-out' growth scenario and produce steep metallicity gradients. However, they do not reproduce the diverse range of metallicity gradients in high-redshift galaxies (only one or the other). In addition, many simulations used empirical feedback models where galactic winds are generated by manually kicking particles and enforcing these wind particles to be temporarily decoupled from hydrodynamics \citep[e.g.][]{dave.2011:cosmo.sim.metal,torrey.2014:illutris.scaling,angles.2014:galactic.outflow} or artificially preventing SNe bubbles from cooling for much longer time \citep[e.g.][]{stinson.2013:magicc.method}. Such models do not properly resolve the launch and propagation of galactic winds from the ISM scale and tend to artificially mix metals on larges scales and prevent strong metallicity gradients from forming. 

In this work, we study the origin and evolution of galaxy metallicity gradients using 32 cosmological zoom-in simulations from the Feedback In Realistic Environments project \citep[FIRE;][]{hopkins.2014:fire.galaxy}\footnote{\href{http://fire.northwestern.edu}{http://fire.northwestern.edu}}. These simulations include physically motivated models of the multi-phase interstellar medium (ISM), star formation, and stellar feedback, with sufficient spatial and mass resolution down to giant molecular cloud (GMC) scales to explicitly resolve the launch and propagation of galactic winds. This is essential in studying metallicity gradients using simulations. In previous studies, it has been shown that these simulations reproduce many observed scaling relations, such as the stellar mass--halo mass relation, the Kennicutt--Schmidt relation, the star-forming main sequence \citep{hopkins.2014:fire.galaxy}, and the MZR \citep{ma.2016:fire.mzr}, for a broad range of halo mass and redshift, {\em without} the need for fine-tuning. These simulations also predict a reasonable covering fraction of neutral absorbers in the circum-galactic medium (CGM) at both low and high redshift \citep{faucher.2015:fire.neutral.hydrogen,faucher.2016:fire.neutral.hydrogen,hafen.2016:fire.lyman.limit}, mass loading factor of galactic outflows \citep{muratov.2015:fire.mass.loading}, and density profiles, kinematics, and chemical abundances of local dwarf galaxies \citep{onorbe.2015:fire.dwarf,chan.2015:fire.cusp.core}, all broadly consistent with observational constraints. All of these demonstrate the validity of using the FIRE simulations to study metallicity gradients. 

Almost all galaxies in the FIRE simulations at high redshift ($z>0.5$) show strong variability (burstiness) in star formation rates (SFRs) on short time-scales of order 10\,Myr \citep{hopkins.2014:fire.galaxy,sparre.2015:fire.sf.burst,muratov.2015:fire.mass.loading,feldmann.2016:massive.fire.long}. In these systems, rapid gas inflows trigger starbursts in the galactic center \citep{torrey.2016:nuclear.starburst}. In turn, feedback from newly formed stars injects sufficient energy and momentum into the ISM to destroy the gas disk and launch galactic winds. At lower redshift ($z<0.5$), on the other hand, massive galaxies ($\Ms\gtrsim10^{10}\,\Msun$) have calmed down, with star formation in the disk being regulated by gas infall and feedback to more stable rates \citep[e.g.][]{faucher.2013:self.regulation}, and feedback can no longer damage the disk nor drive strong gas outflows \citep{muratov.2015:fire.mass.loading}. This transition is likely due to a combination of decreasing galaxy merger rates \citep[e.g.][]{hopkins.2010:merger.rates} and decreasing gas fractions in galaxies \citep[e.g.][]{hayward.hopkins.2015:wind.driving} at low redshift. In this paper, we show that the FIRE simulations reproduce the diversity of kinematics and metallicity gradients observed in high-redshift galaxies. We also show that bursty star formation can produce the observed diversity -- a galaxy may change kinematic properties and metallicity gradient between starburst episodes. This is important for the interpretation of the observed metallicity gradients in high-redshift galaxies.

The paper is organized as follows. We start by introducing the simulations and describing the methods to measure kinematic properties and gas-phase metallicity gradient in the simulated galaxies in Section \ref{sec:method}. We present the main results in Section \ref{sec:results} and discuss and conclude in Section \ref{sec:con}.

We adopt a standard flat $\Lambda$CDM cosmology with cosmological parameters $H_0=70.2 {\rm\,km\,s^{-1}\,Mpc^{-1}}$, $\Omega_{\Lambda}=0.728$, $\Omega_{m}=1-\Omega_{\Lambda}=0.272$, $\Omega_b=0.0455$, $\sigma_8=0.807$ and $n=0.961$, broadly consistent with observations \citep[e.g.][]{hinshaw.2013:wmap9.cosmo.param,planck.2014:cosmo.param}.

\begin{figure*}
\centering
\includegraphics[width=0.93\textwidth]{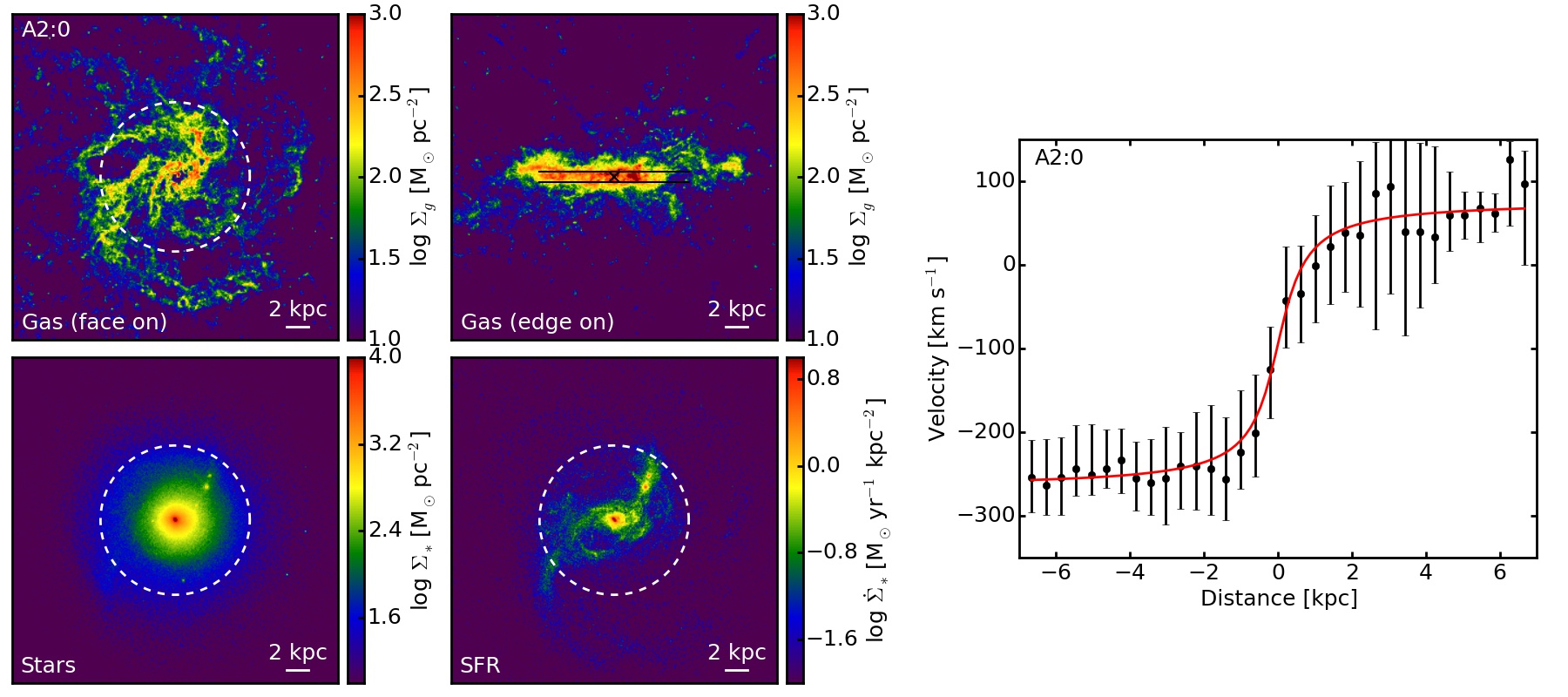} \\
\includegraphics[width=0.93\textwidth]{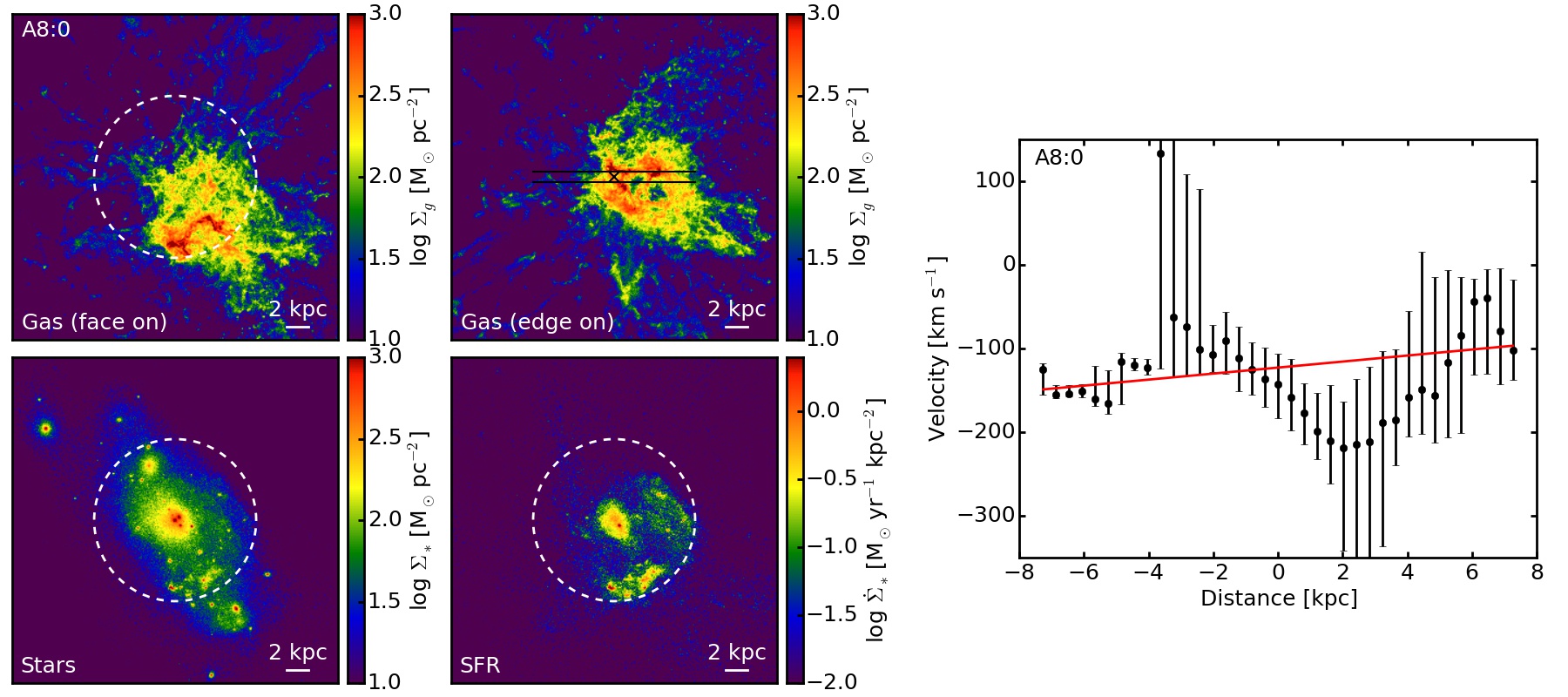} \\
\includegraphics[width=0.93\textwidth]{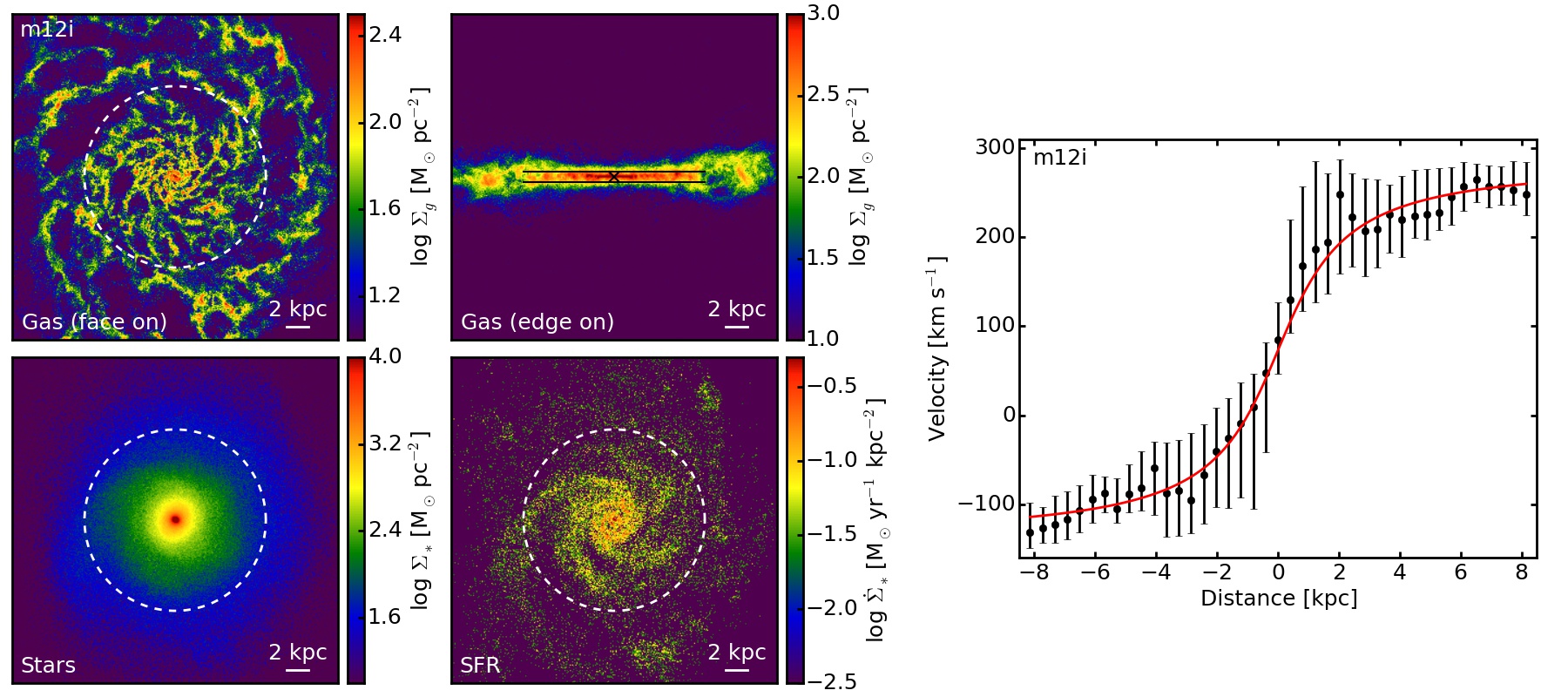}
\caption{{\em Left:} Example images from our simulated sample, including face-on gas surface density (upper left), edge-on gas surface density (upper right), stellar surface density (lower left), and star formation surface density (lower right). We show A2:0 (top) and A8:0 (middle) at $z=2$ and m12i (bottom) at $z=0$ (see Table \ref{tbl:sim} for details). The white circles show $R_{90}$ as defined in Section \ref{sec:kin}. The black lines on the edge-on gas images show the long slits where we extract the gas velocity curve. {\em Right:} Velocity curve extracted from the slit. The symbols and errorbars show the line-of-sight velocity and velocity dispersion, respectively. The red lines show the best fit from the arctan function given by Equation (\ref{eqn:arctan}). A2:0 and m12i have well-ordered rotating disk, while A8:0 is a merging system with no evidence of rotation.}
\label{fig:image}
\end{figure*}

\section{Methodology}
\label{sec:method} 
\subsection{Simulation Details}
\label{sec:sim}
In this work, we use a suite of simulations from the FIRE project that have been presented in previous studies \citep{hopkins.2014:fire.galaxy,faucher.2015:fire.neutral.hydrogen,chan.2015:fire.cusp.core,feldmann.2016:fire.quenching.letter,hafen.2016:fire.lyman.limit}. These are cosmological ``zoom-in'' simulations that are run using {\sc gizmo} \citep{hopkins.2015:gizmo.code} in {\sc p-sph} mode \citep{hopkins.2013:psph.code}. Because of computational expense, some of them are only run to $z=2$, and span a halo mass $10^{11}$--$10^{13}\,\Msun$ at that redshift. For those that are run to $z=0$, we only include the ones above $z=0$ halo mass $10^{11}\,\Msun$ in this study, since smaller galaxies lack observational probes at high redshift. All the simulations used in this paper, along with the mass of the most massive halo in the zoom-in region, the initial mass of baryonic and dark matter particles, minimum force softening lengths, and the reference where the simulation is first presented, are listed in Table \ref{tbl:sim}. We briefly summarize the physical models below for completeness, but refer to \citet[][and references therein]{hopkins.2014:fire.galaxy} for more detailed description.

In our simulations, gas follows an molecular-atomic-ionized cooling curve from 10--$10^{10}$\,K, including metallicity-dependent fine-structure and molecular cooling at low temperatures and high-temperature metal-line cooling followed species-by-species for 11 separately tracked species \citep[H, He, C, N, O, Ne, Mg, Si, S, Ca, and Fe; see][]{wiersma.2009:metal.cooling}. At each timestep, the ionization states and cooling rates are determined following \citet{katz.1996:sph.cooling} for primordial abundances and from a compilation of {\sc cloudy} runs for metals, including a uniform but redshift-dependent photo-ionizing background tabulated in \citet{faucher.2009:uvb}, and photo-ionizing and photo-electric heating from local sources. Gas self-shielding is accounted for with a local Jeans-length approximation.

We allow star formation to take place only in dense, molecular, and self-gravitating regions with hydrogen number density above a threshold $\nc=5$--$50\,\cm^{-3}$ \citep{hopkins.2013:sf.criteria}. Stars form at 100\% efficiency per local free-fall time when the gas meets these criteria and there is no star formation elsewhere. A star particle inherits the metallicity of each tracked species from its parent gas particle. Every star particle is treated as a single stellar population with known mass, age, and metallicity, assuming a \citet{kroupa.2002:imf} initial mass function (IMF) from $0.1$--$100\,\Msun$. Then the ionizing photon budgets, luminosities, Type II supernova rates, mechanical luminosities of stellar winds, etc., are directly tabulated from the stellar population models in {\sc starburst99} \citep{leitherer.1999:sb99}. The Type Ia SN rates follow the time delay distribution from \citet{mannucci.2006:snia.rates}. We account for the following stellar feedback mechanisms, including (1) local and long-range momentum flux from radiative pressure, (2) energy, momentum, mass and metal injection from SNe and stellar winds, and (3) photo-ionization and photo-electric heating. We follow \citet{wiersma.2009:chemical.enrich} and account for metal production from Type-II SNe, Type-Ia SNe, and stellar winds using the metal yields in \citet{woosley.weaver.1995:sneii.yield}, \citet{iwamoto.1999:snia.yield}, and \citet{izzard.2004:agb.yield}, respectively. We do not include a sub-resolution metal diffusion model, but the simulations explicitly resolve the metal mixing by advection of gas particles.

\subsection{Galaxy Identification and Definitions}
\label{sec:galaxy}
We use Amiga's Halo Finder \citep[{\sc ahf};][]{knollmann.knebe.2009:ahf.code} to identify halos in the simulations. The approximate halo mass at $z=2$ and $z=0$ (if applicable) for the most massive (best-resolved) halo in each simulation are listed in Table \ref{tbl:sim}, where we adopt the redshift-dependent virial parameter from \citet{bryan.norman.1998:xray.cluster}. In this paper, we only study the central galaxy in the most massive halo in each simulation. The entire simulated sample is only studied at four redshifts $z=2$, 1.4, 0.8, and 0 (if applicable). The physical properties of these galaxies (as described below) at these redshifts are presented in Appendix \ref{sec:append:gal}.

We define the center of each galaxy by iteratively finding the geometric center of all star particles within a sphere of decreasing radius from 20\,kpc to 1\,kpc. This generally corresponds closely to the location of maximum stellar mass density. The stellar mass ($\Ms$) and the star formation rate (SFR) for the central galaxy are measured within 10\,kpc from this center, where we remove the contamination of satellite galaxies if necessary. The SFRs are averaged over 200\,Myr to mimic the observational measurements based on far-ultraviolet luminosity \citep[e.g.][]{sparre.2015:fire.sf.burst}. Next, we define a characteristic radius $R_{90}$, which encloses 90\% of the star formation within 10\,kpc. Such definition of galactic center and characteristic radius appears to be most numerically stable, given that a considerable fraction of galaxies in our simulated sample have clumpy and irregular morphologies (especially those at high redshifts). The stellar mass, SFR, and $R_{90}$ for the entire simulated sample are listed in Appendix \ref{sec:append:gal}. Our sample covers a stellar mass range $10^8$--$10^{11}\,\Msun$.

For simplicity, we define the $z$-axis to be aligned with the total angular momentum of all gas particles within $R_{90}$ and the $x$-axis to be an arbitrary direction perpendicular to $z$-axis. We refer to face-on and edge-on views when observing along the $z$- and $x$-axis, respectively. In Fig. \ref{fig:image} (left two columns), we show example images for three galaxies in our sample, A2:0 at $z=2$ (top), A8:0 at $z=2$ (middle), and m12i at $z=0$ (bottom). For each galaxy, we show a face-on gas image ($x$-$y$ plane, top left) and edge-on gas image ($y$-$z$ plane, top right), face-on stellar image (bottom left), and face-on SFR map (bottom right, averaged over 200\,Myr). The dashed white circles on all face-on images show the characteristic $R_{90}$ of each galaxy. A8:0 is a merging system that has clumpy, irregular morphology, while A2:0 and m12i have star-forming gas disks.

\begin{figure}
\centering
\includegraphics[width=\linewidth]{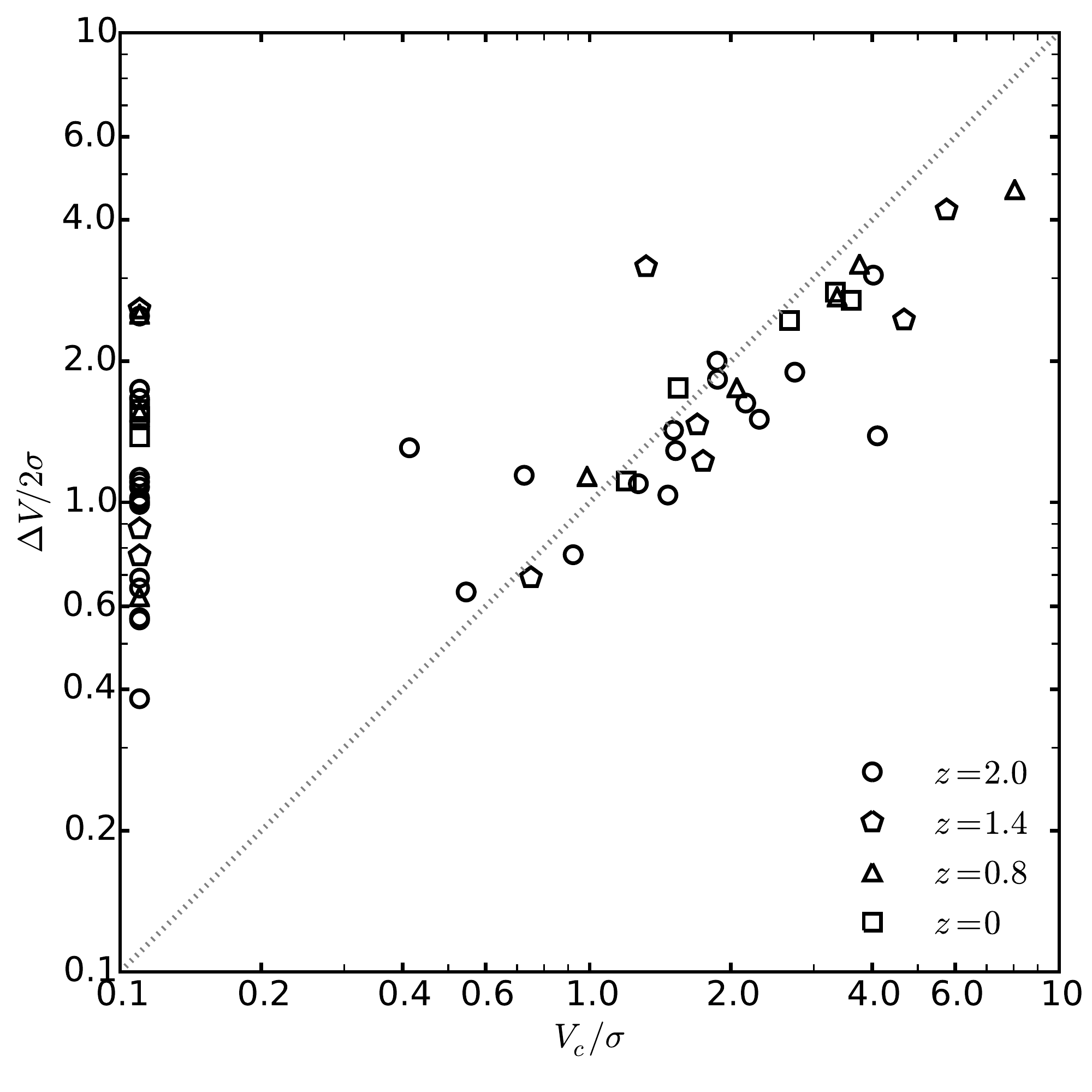}
\caption{Comparison between $V_c/\sigma$ and $\Delta V/2\sigma$ for our simulated galaxies. $V_c$ is the rotation velocity given by the best fit of the velocity curve by the arctan function in Equation (\ref{eqn:arctan}), while $\Delta V$ is the peak-to-peak velocity difference. $\sigma$ is the maximum velocity dispersion (see Figure \ref{fig:image} for examples). Galaxies that cannot be well fitted by an arctan function are plotted at $V_c/\sigma\sim0.1$. $V_c/\sigma$ and $\Delta V/2\sigma$ are broadly consistent with each other for galaxies with $V_c/\sigma\geq1$, indicating that they have well-ordered rotation by either criterion. However, galaxies with $V_c/\sigma<1$ show $\Delta V/2\sigma\sim0.4$--3. This suggests that $\Delta V/2\sigma$ is ambiguous for non-rotationally supported systems.}
\label{fig:vsigma}
\end{figure}

\begin{figure*}
\centering
\begin{tabular}{ccc}
\centering
  \includegraphics[width=0.31\textwidth]{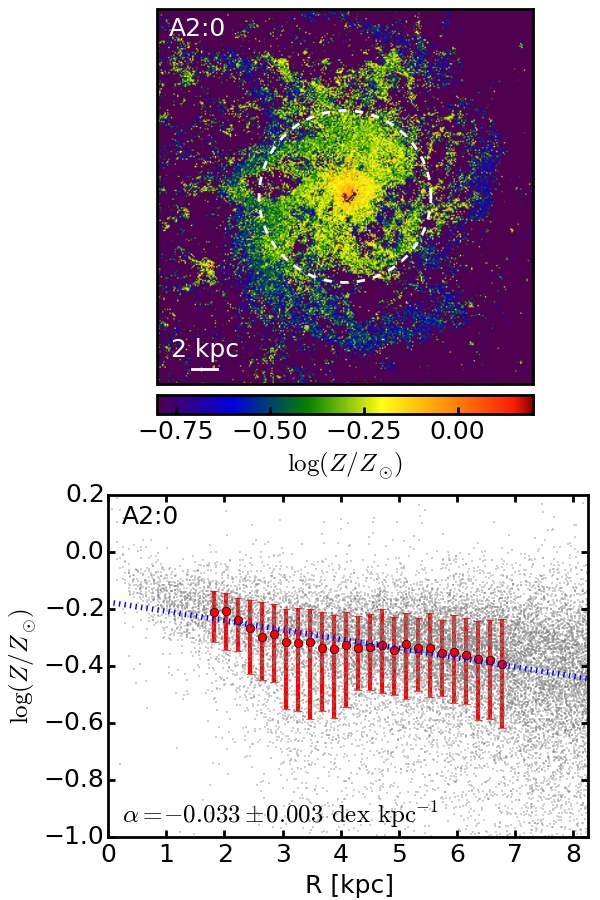} & 
  \includegraphics[width=0.31\textwidth]{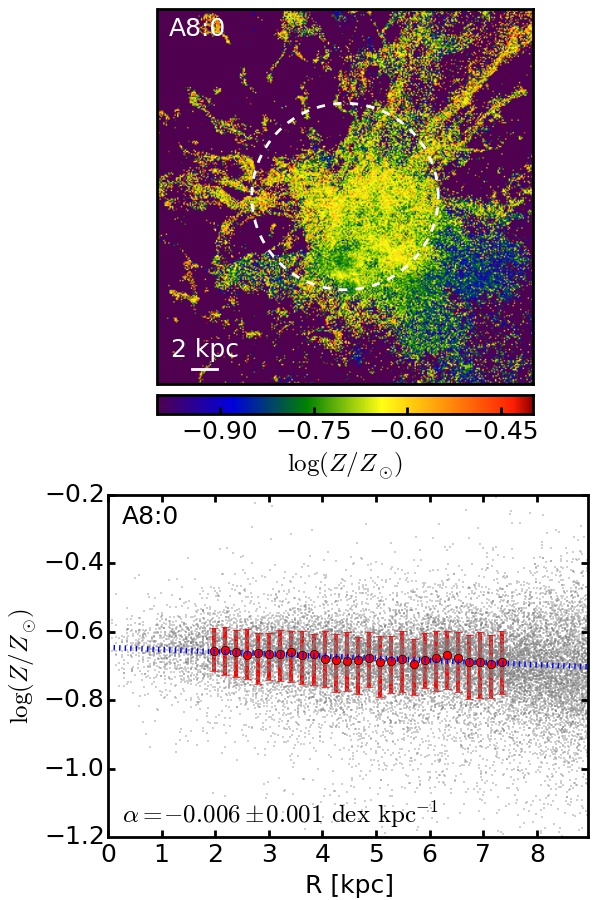} & 
  \includegraphics[width=0.31\textwidth]{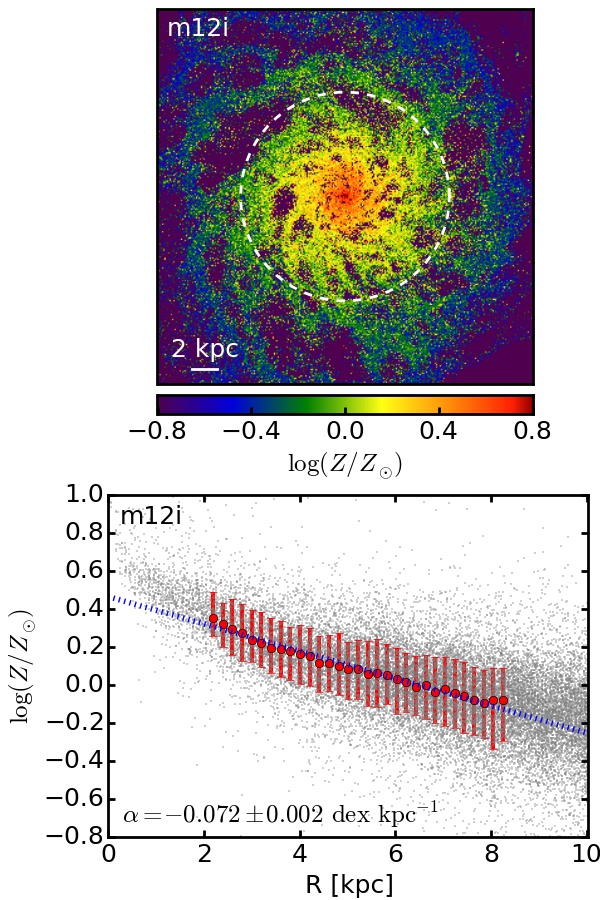}
\end{tabular}
\caption{{\em Top:} Face-on gas-phase metallicity map for the three example galaxies in Fig. \ref{fig:image}. {\em Bottom:} Metallicity profile. The grey points show individual pixels, and the red points and errorbars show the median and $1\sigma$ dispersion of metallicity at every 0.2\,kpc in 0.25--$1R_{90}$. The blue lines show the best linear fit $\log(Z/\Zsun)=\alpha R+\beta$, where $\alpha$ gives the slope of metallicity gradient in $\dkpc$.}
\label{fig:rgrad}
\end{figure*}

\subsection{Kinematics}
\label{sec:kin}
Before we present the gas-phase metallicity gradients for our simulated sample, we first measure the kinematic properties of these galaxies, as commonly done in observational studies \citep[e.g.][]{yuan.2011:metal.gradient.spiral,jones.2013:lense.metal.grad,leeth.2016:lense.metal.grad}. We do so by mimicking the widely-used long-slit spectroscopy technique. The mock slit is placed along the $y$-axis (edge-on) along the mid-plane with a vertical width of 1\,kpc, as illustrated by the black lines on the edge-on gas images in Fig. \ref{fig:image}. We then extract the one-dimensional velocity curve along the slit. We measure the line-of-sight gas velocity and $1\sigma$ velocity dispersion in the range $-R_{90}<y<R_{90}$ with a spatial resolution of $\Delta y=0.4\,\kpc$, by taking into account all gas particles with number density $n>1\,\cm^{-3}$ in every pixel. This allows us to primarily select interstellar gas and eliminate contamination by foreground/background gas in the circumgalactic/intergalactic medium. Example velocity curves of the three galaxies, A2:0, A8:0 (at $z=2$), and m12i (at $z=0$), are shown in the right column of Fig. \ref{fig:image}, with the black points and errorbars representing the line-of-sight velocity and velocity dispersion along the slit.

We fit the one-dimensional velocity curve with the following analytic form
\be
\label{eqn:arctan}
V(R) = V_0 + V_c \frac{2}{\pi} \arctan \left( \frac{R}{R_t} \right),
\ee
as motivated by the simple disk model commonly adopted in various studies \citep[e.g.][]{jones.2010:zgrad.z2.inside.out,swinbank.2012:hizels.metal.grad,stott.2014:metal.grad.z1.kmos,leeth.2016:lense.metal.grad}. For our simulated galaxies, $V_0$ accounts for the peculiar velocity in the simulation box and $V_c$ gives the asymptotic circular velocity at large radii. Example fits for the three galaxies are shown by the red lines in Fig. \ref{fig:image}. The velocity curves of A2:0 and m12i can be well described by the arctan function, reaffirming that these galaxies have well-ordered rotating disks. However, the chaotic system, A8:0, returns a bad fit (as reflected by unphysical values of $V_c$). We have visually checked all of our simulations and find that bad fits occur when a galaxy has clumpy, irregular morphology and shows little evidence of rotation. For these galaxies, $V_c$ cannot be properly defined. We also follow \citet{leeth.2016:lense.metal.grad} and measure the ``peak-to-peak'' velocity difference $\Delta V$ along the slit. Any galaxy can give a finite $\Delta V$ despite its kinematic properties. For a rotating disk, $\Delta V$ equals $2V_c$ in the asymptotic limit and is thus a proxy for the rotation velocity. We define the velocity dispersion of the galaxy $\sigma$ as the maximum velocity dispersion along the slit. $V_c$, $\Delta V$, and $\sigma$ for the entire simulated sample are presented in Appendix \ref{sec:append:gal}. Note that some galaxies in our sample are temporarily quenched, with little gas in the central region. The kinematic properties for these galaxies cannot be properly determined.

The degree of rotational support of a galaxy can be defined as either $V_c/\sigma$ or $\Delta V/2\sigma$. In Fig. \ref{fig:vsigma}, we compare $V_c/\sigma$ and $\Delta V/2\sigma$ for our simulated galaxies. For illustrative purposes, we plot those whose velocity curve cannot be well fitted by Equation \ref{eqn:arctan} at $V_c/\sigma\sim0.1$, as they do not have well-ordered rotation. The criterion for rotationally supported system is commonly taken to be $V_c/\sigma\geq1$ or $\Delta V/2\sigma\geq0.4$ \citep[e.g.][]{forster.2009:sins.survey.ifu.z2,leeth.2016:lense.metal.grad}. Most of our simulated galaxies with $V_c/\sigma\geq1$ have consistent values of $\Delta V/2\sigma$, reaffirming that these galaxies are rotationally supported. However, galaxies with $V_c/\sigma<1$ span a wide range of $\Delta V/2\sigma$, mostly from 0.4--3 for our simulated sample. These galaxies have little evidence of rotation as shown by the velocity curve and confirmed by our visual inspection, but they would be classified as rotating systems by the criterion $\Delta V/2\sigma\geq0.4$. We caution that $\Delta V/2\sigma$ is an ambiguous indicator in practice, especially for those galaxies with little rotation.

\begin{figure}
\centering
\includegraphics[width=\linewidth]{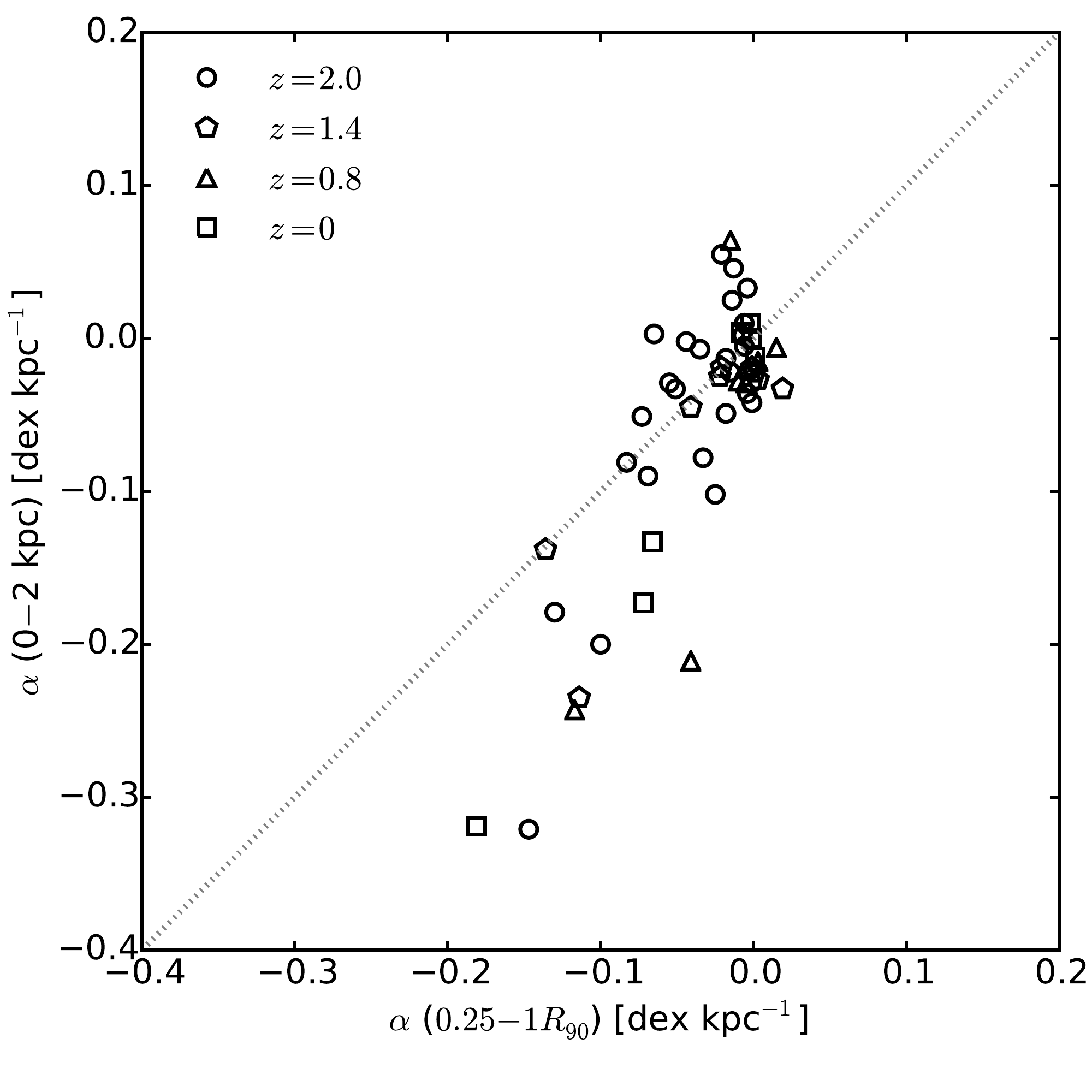}
\caption{Metallicity gradients measured in the radial interval 0.25--$1R_{90}$ vs. metallicity gradients measured in the central 0--2\,kpc. The difference is small when the gradient is flat, because the gas-phase metallicity is almost uniform in the ISM (e.g. simulation A8:0). On the other hand, the slopes measured over 0--2\,kpc are much steeper than those measured over 0.25--$1R_{90}$ in galaxies that show strong negative metallicity gradients (e.g. simulations A2:0 and m12i shown in Figs. \ref{fig:image} and \ref{fig:rgrad}). They show rapidly increasing metallicity profiles toward the galactic center.}
\label{fig:a1a2}
\end{figure}

\subsection{Metallicity Gradients}
\label{sec:grad}
We now present the metallicity gradients for our simulated sample. In the top panel of Fig. \ref{fig:rgrad}, we show the face-on metallicity map for the same galaxies as in Fig. \ref{fig:image}, with a pixel size of 100\,pc. We measure the mass-weighted metallicity of all gas particles in each pixel. We only show pixels where the gas surface density is above $\Sigma_g\geq10\,\Msun\,\pc^{-2}$. Such threshold is roughly the surface density above which fragmentation and star formation occurs in these simulations and observations (M. Orr et al., in preparation), so these pixels are likely to have observationally detectable nebular emission lines. This also reduces the shot noise from low surface density pixels where the metallicities are determined by individual gas particles. In the bottom panels, we plot the gas-phase metallicity as a function of projected radius for individual pixels (grey points). Again, only pixels above surface density $10\,\Msun\,\pc^{-2}$ are shown. We measure the median metallicity and its $1\sigma$ dispersion at every 0.2\,kpc in a certain radius interval, as illustrated by the red points and errorbars (in 0.25--$1R_{90}$, our fiducial interval) in Fig. \ref{fig:rgrad}. We require a minimum number of 20 pixels in a 0.2\,kpc bin to obtain a reliable measurement at this radius. We fit the metallicity profile with a linear function (the blue dotted lines in Fig. \ref{fig:rgrad})
\be
\label{eqn:zgrad}
\log(Z/\Zsun) = \alpha R + \beta, 
\ee
to obtain the slope of the metallicity gradient $\alpha$ (in $\dkpc$).

Equation \ref{eqn:zgrad} is motivated by the fact that metallicity gradients are most commonly measured in $\dd\log Z/\dd R$ (in $\dkpc$) in the literature, although the metallicity profile of a galaxy may deviate from a linear function in reality. In Fig. \ref{fig:a1a2}, we compare the slopes of the metallicity gradients measured over 0.25--$1R_{90}$ and over 0--2\,kpc, respectively. Both slopes are qualitatively consistent with each other. The difference is small when the gradient is close to flat, because the metals are nearly uniformly mixed within the ISM (e.g. simulation A8:0). On the other hand, galaxies with strong negative metallicity gradients tend to have a rapidly increasing metallicity profile toward the center (e.g. simulations A2:0 and m12i in Figs. \ref{fig:image} and \ref{fig:rgrad}), as reflected by the fact that the slopes measured in 0--2\,kpc are much steeper than those measured in 0.25--$1R_{90}$. This happens in our simulations because the galactic centers can reach very high gas surface densities ($\Sigma_g\gtrsim10^3\,\Msun\,\pc^{-2}$) during a starburst, and the star formation efficiency may increase dramatically with gas surface density (e.g. \citealt{burkert.hartmann.2013:sf.efficiency,torrey.2016:nuclear.starburst}; M. Grudi{\'c} et al., in preparation), resulting in rapid metal enrichment toward the center. Such picture is consistent with previous studies on the formation of cusp elliptical galaxies via mergers \citep[e.g.][]{hopkins.2009:cusp.ellipticals}, which reproduce the observed steep metallicity gradients in the central region of early-type galaxies \citep[e.g.][]{reda.2007:resolve.early.type,sanchez.2007:resolve.early.type}. 
In this work, we primarily focus on the metallicity gradients measured over 0.25--$1R_{90}$. The slopes of metallicity gradient for the entire simulated sample are listed in Appendix \ref{sec:append:gal}. We note that all of our results presented below are qualitatively consistent if one uses the gradients measured in 0--2\,kpc. A detailed discussion on the full metallicity profile is beyond the scope of this study, but worth further investigations in future work.

\section{Results}
\label{sec:results}
\subsection{Metallicity gradients: general properties}
\label{sec:general}
As illustrated by the visual examples in Fig. \ref{fig:rgrad} and more quantitative results shown in Appendix \ref{sec:append:gal}, our simulations produce a variety of kinematic properties and metallicity distributions. Simulations A2:0 and m12i have obvious negative metallicity gradients, with the center being more metal-enriched than the outskirts, consistent with the observed metallicity patterns in local and some high-redshift galaxies \citep[e.g.][]{zaritsky.1994:metal.grad.spiral,vanzee.1998:metal.grad.spiral,yuan.2011:metal.gradient.spiral,jones.2013:lense.metal.grad,sanchez.2014:califa.metal.grad}. Both of them have a rotationally supported, star-forming disk as shown in Fig. \ref{fig:image}. In contrast, simulation A8:0 is a merging system that has a clumpy, irregular gas morphology with no well-ordered gas motion, and a relatively uniform metallicity distribution, with metallicity gradient close to flat. Intuitively, these examples indicate that strong negative metallicity gradients are more likely to occur in galaxies with a rotating disk, while strongly perturbed galaxies tend to have flat gradients. 

Strong perturbations, mostly induced by mergers, rapid gas infall, and strong outflows, can stir the gas and drive galactic-scale motion in the ISM, with typical velocities up to several hundred km\,s$^{-1}$. This causes gas/metal re-distribution on galactic scales of $\lesssim10\,\kpc$ on relatively short time-scales $\sim 10$--50\,Myr, leading to kinematically hot gas motion and flat metallicity gradients\footnote{Here we do not consider metal mixing on scales below our resolution limit, but rather focus on re-distribution of metals driven by largest-scale motion. 
This is justified by more detailed studies of diffusion processes in supersonically turbulent media like the ISM, which show that diffusion is most efficient on large scales \citep[e.g.][]{colbrook.2016:turbulent.diffusion}.}. In simulation A8:0, the perturbation is induced by a series of minor mergers (see Fig. \ref{fig:image}). Besides, strong stellar feedback can also drive galaxy-scale motion in the ISM, resulting in irregular gas motion and morphology \citep[e.g.][]{agertz.2016:impact.feedback}. \citet{gibson.2013:metal.grad.notes} show that simulations with strong feedback produce flat metallicity gradients, while those with weak feedback tend to produce steep gradients. The high resolution and physically motivated models of stellar feedback adopted in the FIRE simulations enable us to explicitly resolve the launch and propagation of galactic winds from small scales (tens of pc) to galactic scales, which is essential to study gas-phase metallicity gradients. 

The rest of this section is organized as follows. Before going into details about metallicity gradients in our simulated galaxies, we first show where our sample lies on the galaxy MZR in Section \ref{sec:mzr}. In Section \ref{sec:smass}, we will study the dependence of metallicity gradient on stellar mass and specific star formation rate (sSFR). In Section \ref{sec:zkin}, we will examine the relation between metallicity gradient and the degree of rotational support. In Section \ref{sec:zred}, we will present the redshift dependence on metallicity gradient. In Section \ref{sec:feedback}, we will perform a case study on simulation m12i and explore how stellar feedback can change metallicity gradients on short time-scales ($\lesssim\Gyr$), which has a great effect on the interpretations of the observed metallicity gradients in high-redshift galaxies.

\begin{figure}
\centering
\includegraphics[width=\linewidth]{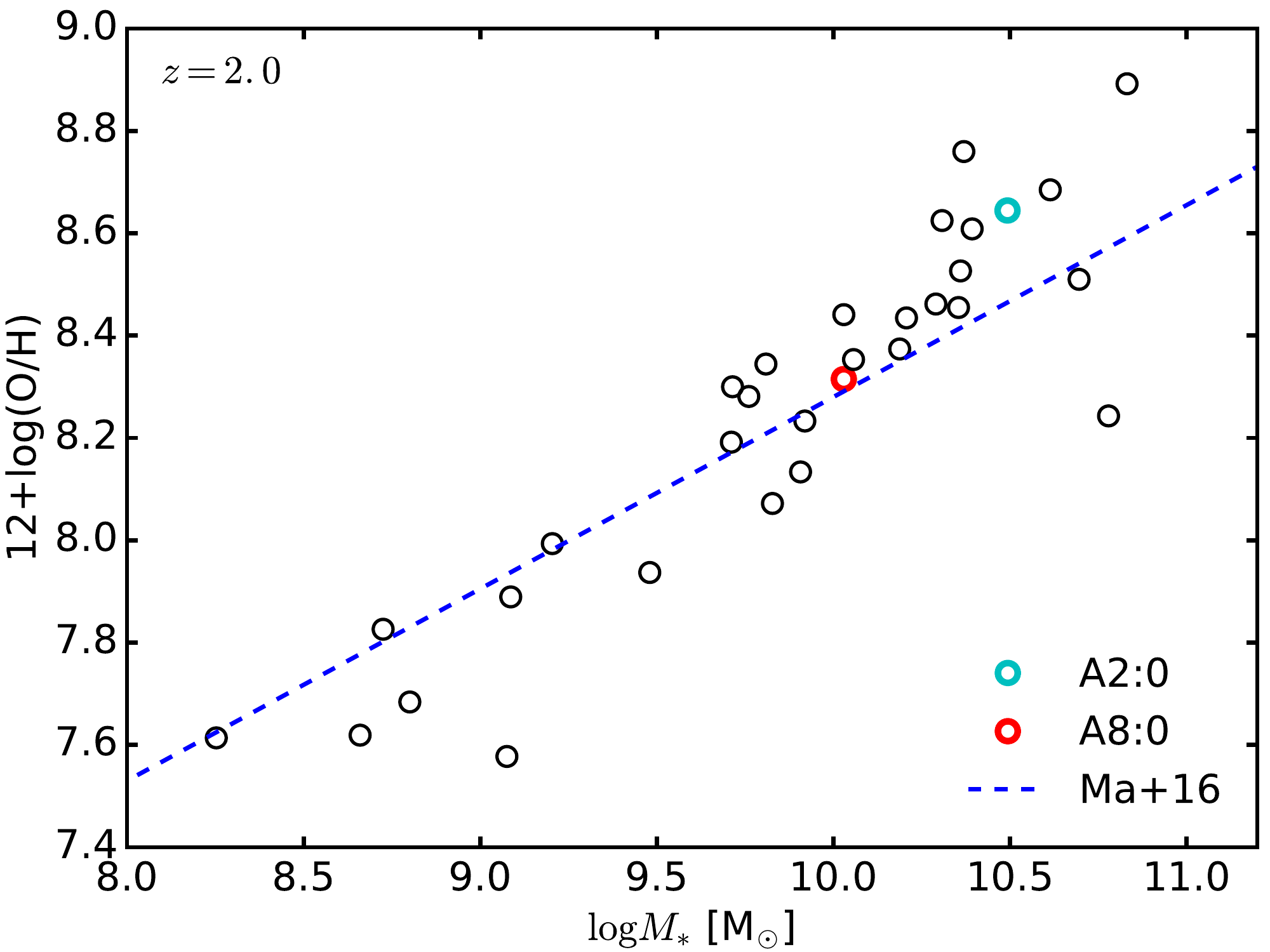}
\caption{Gas-phase oxygen abundance vs. stellar mass for our simulated sample at $z=2$. Galaxies A2:0 and A8:0 (see also Figs. \ref{fig:image} and \ref{fig:rgrad}) are indicated by the thick cyan and red circles, respectively. The simulations analyzed in this work cover a stellar mass range $10^8$--$10^{11}\,\Msun$. The blue dashed line shows the fit from \citet{ma.2016:fire.mzr}, which is derived from a sample covering a stellar mass range $10^4$--$10^{10}\,\Msun$ at this redshift.}
\label{fig:mzr}
\end{figure}

\begin{figure*}
\centering
\begin{tabular}{cc}
\centering
  \includegraphics[width=0.48\textwidth]{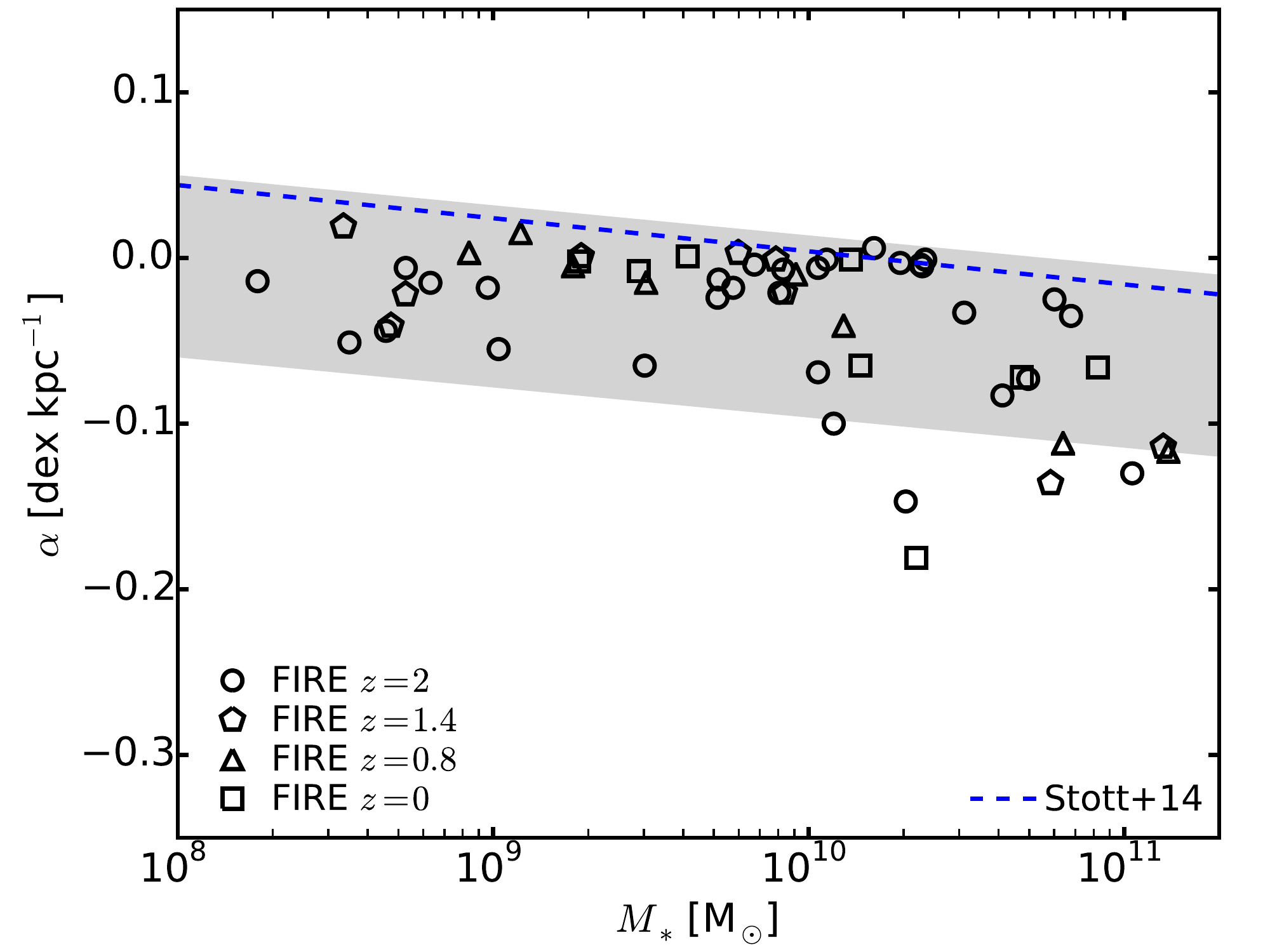} & 
  \includegraphics[width=0.48\textwidth]{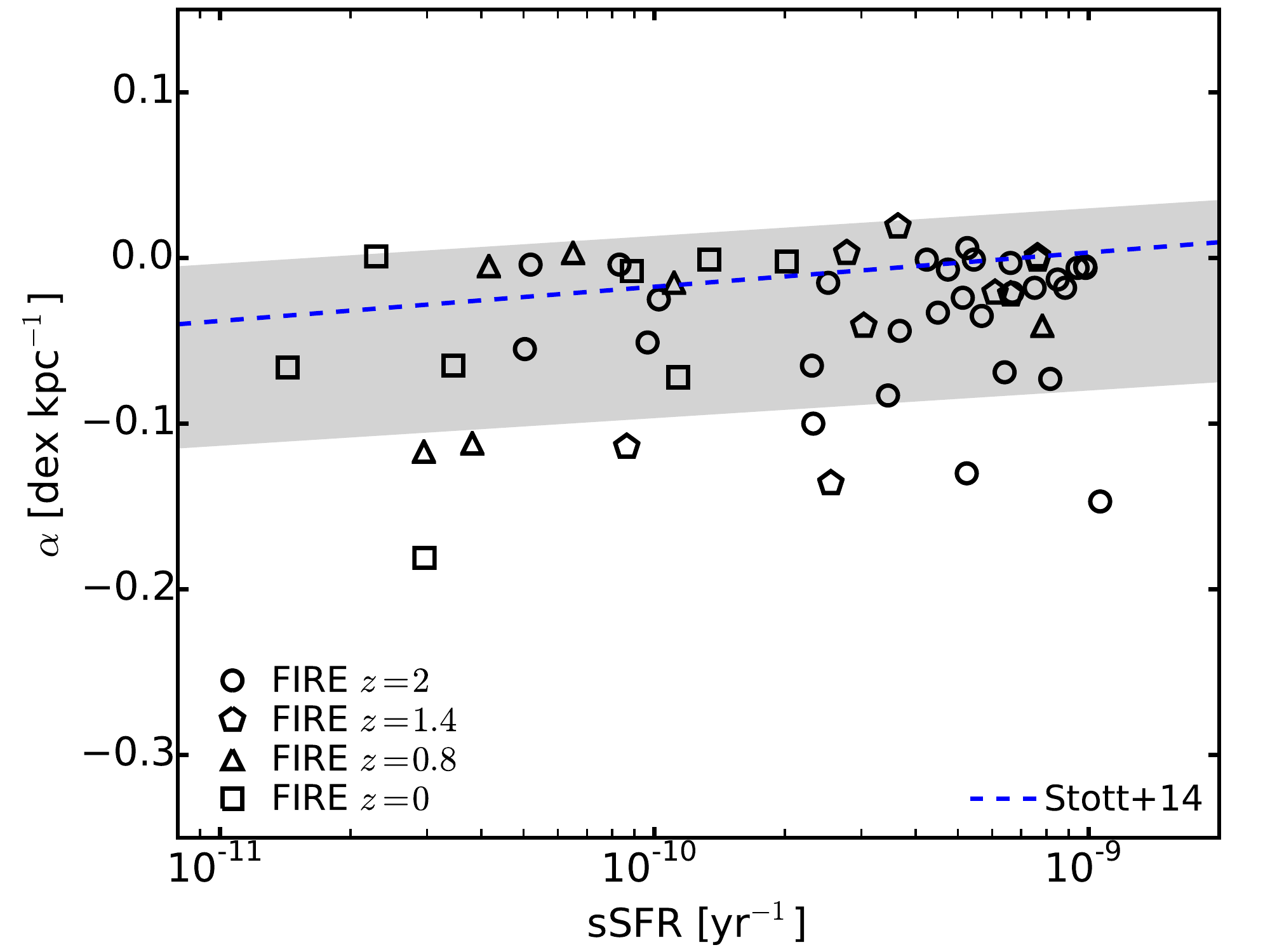} 
\end{tabular}
\caption{{\em Left:} Metallicity gradient (measured over 0.25--$1R_{90}$) vs. stellar mass. {\em Right:} Metallicity gradient vs sSFR. The shaded regions show the $2\sigma$ interval of the linear fit to the simulated data. The blue dashed lines show the linear fit to a compilation of observational data at $z=0$--2.5 from \citet{stott.2014:metal.grad.z1.kmos}. There is weak dependence of metallicity gradient on both stellar mass and sSFR. Low-mass galaxies or those with high sSFR tend to have flat metallicity gradients, due to the fact that feedback is more efficient in such galaxies.}
\label{fig:alms}
\end{figure*}

\begin{figure*}
\centering
\begin{tabular}{cc}
\centering
  \includegraphics[width=0.48\textwidth]{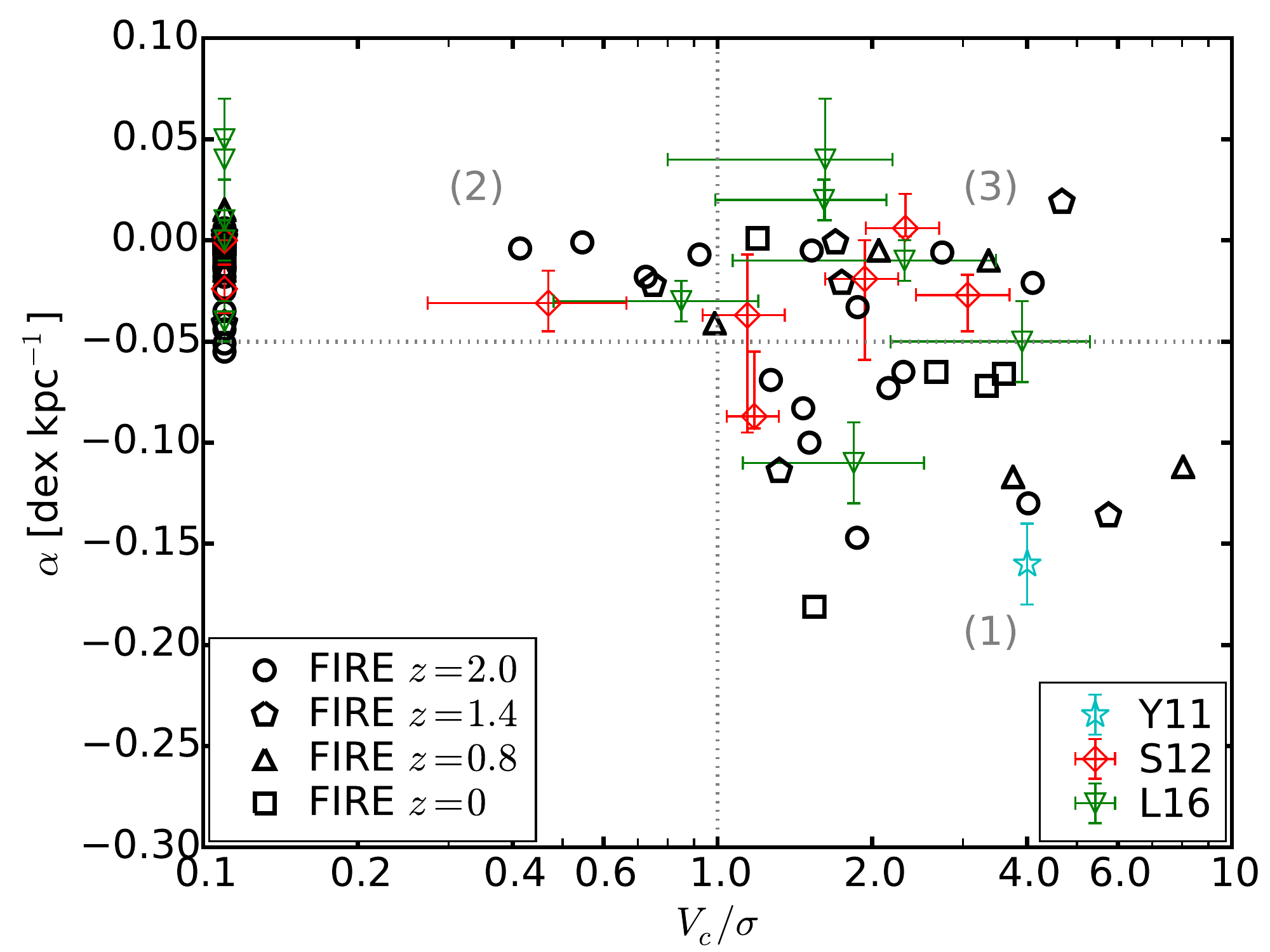} & 
  \includegraphics[width=0.48\textwidth]{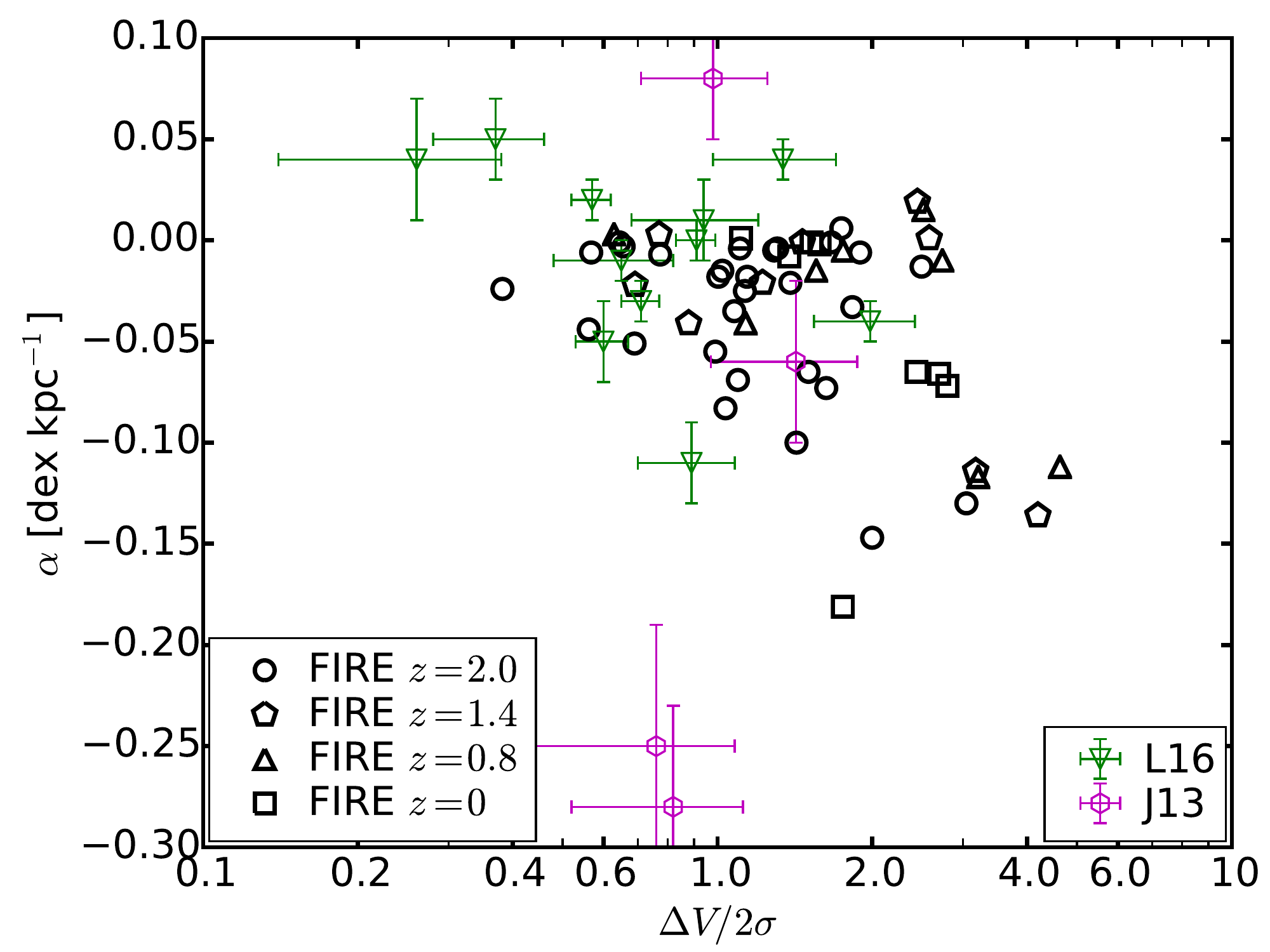} 
\end{tabular}
\caption{Metallicity gradient (measured over 0.25--$1R_{90}$) vs. degree of rotational support. {\em Left:} $\alpha$--$V_c/\sigma$. As in Fig. \ref{fig:vsigma}, galaxies for which $V_c$ cannot be properly determined are plotted at $V_c/\sigma\sim0.1$. The simulated sample can be divided into three populations: (1) strong negative metallicity gradients {\em only} appear in rotationally supported galaxies ($V_c/\sigma\geq1$), (2) highly perturbed galaxies with no rotation ($V_c/\sigma<1$) tend to have flat gradients, and (3) there is also a population of rotationally supported galaxies that have flat metallicity gradients. We do {\em not} find any strongly perturbed galaxy that has a strong negative metallicity gradient. {\em Right:} $\alpha$--$\Delta V/2\sigma$. Similarly, strong negative gradients {\em only} occur in galaxies with $\Delta V/2\sigma\geq1$. Symbols with errorbars show observational data from \citet[][Y11]{yuan.2011:metal.gradient.spiral}, \citet[][S12]{swinbank.2012:hizels.metal.grad}, \citet[][J13]{jones.2013:lense.metal.grad}, and \citet[][L16]{leeth.2016:lense.metal.grad}. Our simulations reproduce the observed complexity in metallicity gradient and kinematic properties.}
\label{fig:alvc}
\end{figure*}

\subsection{The mass-metallicity relation (MZR)}
\label{sec:mzr}
We follow \citet{ma.2016:fire.mzr} and define the gas-phase metallicity as mass-weighted mean metallicity of all gas particles below $10^4$\,K in the central galaxy (satellites excluded). In Fig. \ref{fig:mzr}, we show the gas-phase MZR for our simulated sample at $z=2$, where we define the oxygen abundance as ${\rm 12+log(O/H)}=\log(Z/\Zsun)+9.0$. Galaxies A2:0 and A8:0 shown in Figs. \ref{fig:image} and \ref{fig:rgrad} are indicated with thick cyan and red circles, respectively. They have typical gas-phase metallicities for our sample. 
In \citet{ma.2016:fire.mzr}, we extensively studied the MZR in a sample of FIRE simulated galaxies at $z=1.4$, 0.8, and 0. In that work, we showed that m12i lies on the observed median gas-phase and stellar MZR from \citet{tremonti.2004:gas.mzr.sdss} and \citet{gallazzi.2005:stellar.mzr.sdss} at $z=0$.
The blue dashed line shows the linear fit to the simulations from \citet{ma.2016:fire.mzr}. We note that \citet{ma.2016:fire.mzr} used a sample that covered the stellar mass range from $10^4$--$10^{10}\,\Msun$ at $z=2$, while the new simulations from \citet{feldmann.2016:fire.quenching.letter} included in this work allow us to extend our analysis to $10^{11}\,\Msun$.

\subsection{Metallicity gradient vs stellar mass and sSFR}
\label{sec:smass}
We start by examining the correlation between gas-phase metallicity gradient (measured over 0.25--$1R_{90}$) and galaxy properties. In Fig. \ref{fig:alms}, we show the dependence of metallicity gradient on stellar mass (left) and specific star formation rate (sSFR, right) for the simulated sample at four redshifts $z=2.0$, 1.4, 0.8, and 0. We do not find significant differences between redshifts except perhaps for massive galaxies at $z\sim0$, consistent with recent observations \citep[e.g.][]{wuyts.2016:kmos3d.metal.grad}. The shaded regions show $2\sigma$ linear fits to the simulated data. We find a weak anti-correlation between metallicity gradient and stellar mass. Low-mass galaxies tend to have flat gradients, because feedback is very efficient in driving outflows and thus mixing metals in low-mass systems \citep{muratov.2015:fire.mass.loading,muratov.2016:fire.metal.loading}. Note that the FIRE project also includes simulations of isolated dwarf galaxies with stellar masses $\Ms\sim10^4$--$10^8\,\Msun$ \citep[e.g.][]{hopkins.2014:fire.galaxy,chan.2015:fire.cusp.core}, but we do not consider these dwarfs in this work, because observations probe only galaxies more massive than $10^9\,\Msun$. Nevertheless, they also have very weak (flat) metallicity gradients \citep{elbadry.2016:fire.migration}, because they are bursty, feedback-dominated galaxies, consistent with the argument above. We also find a weak correlation between metallicity and sSFR. Most galaxies with high sSFR have undergone rapid gas inflows that trigger starbursts, and feedback in turn drives strong outflows. Such violent gas infall and outflows can stir the gas in the ISM and mix metals on galactic scales efficiently, resulting in a flat metallicity gradient. In Fig. \ref{fig:alms}, we also show the linear fits to a compilation of observational data at redshifts $z=0$--2.5 from \citet[][blue dashed lines]{stott.2014:metal.grad.z1.kmos}. These trends are in qualitative agreement with our simulations, but we note that both observations and our simulations only show weak trends with stellar mass and sSFR (within $3\sigma$, the data are consistent with no trend).

\begin{figure}
\centering
\includegraphics[width=\linewidth]{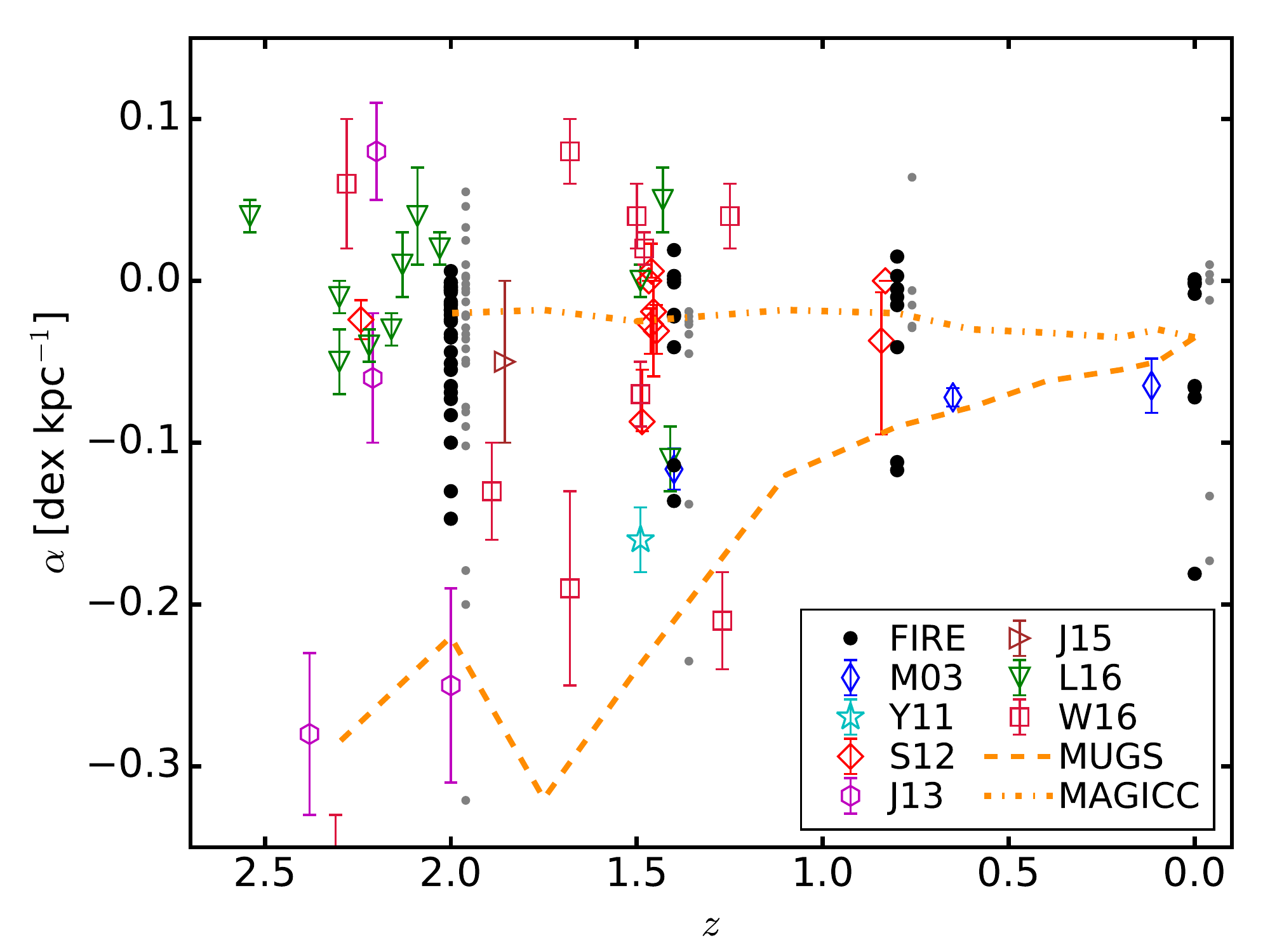}
\caption{Metallicity gradient vs. redshift. The black points show the metallicity gradients measured in 0.25--$1R_{90}$ for the entire FIRE sample at four redshifts. The smaller grey points show the slopes measured in 0--2\,kpc. The grey points are shifted slightly right along the $x$-axis for better illustration. Symbols with errorbars show a compilation of observations from \citet[][M03]{maciel.2003:mw.metal.grad.evolve}, \citet[][Y11]{yuan.2011:metal.gradient.spiral}, \citet[][S12]{swinbank.2012:hizels.metal.grad}, \citet[][J13]{jones.2013:lense.metal.grad}, \citet[][J15]{jones.2015:glass.metal.grad}, \citet[][L16]{leeth.2016:lense.metal.grad}, and \citet[][W16]{wang.2016:glass.metal.grad}. The green lines show the predictions from the sub-grid `conservative' (weak) feedback model used in MUGS simulations (dashed) and the `enhanced' (strong) feedback used in MAGICC simulations (dotted) from \citet{gibson.2013:metal.grad.notes}. Our simulations agree well with the wide range of metallicity gradients observed over the $z=0$--2.5 redshift range -- in some circumstances (e.g. starbursts), feedback is predicted to be effectively `strong' to produce flatten metallicity gradients, while in others, it is sufficiently `weak' to allow a strong negative gradient.}
\label{fig:alz}
\end{figure}

\begin{figure*}
\centering
\begin{tabular}{cc}
\centering
  \includegraphics[width=\linewidth]{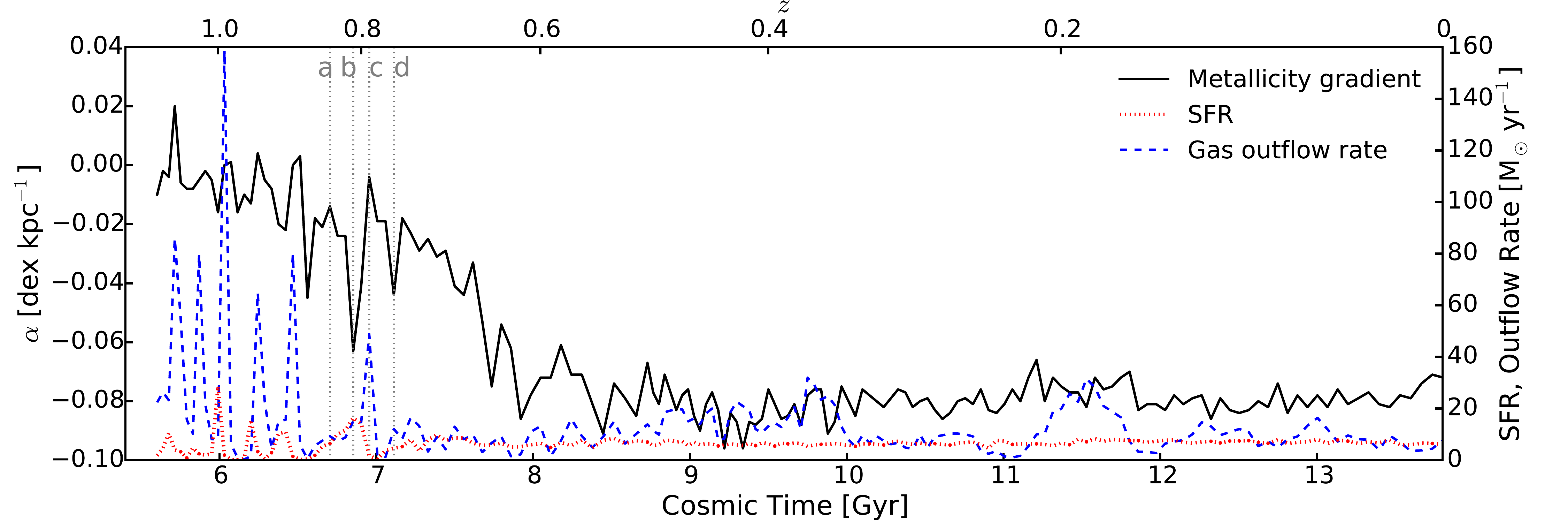} & \\
  \includegraphics[width=\linewidth]{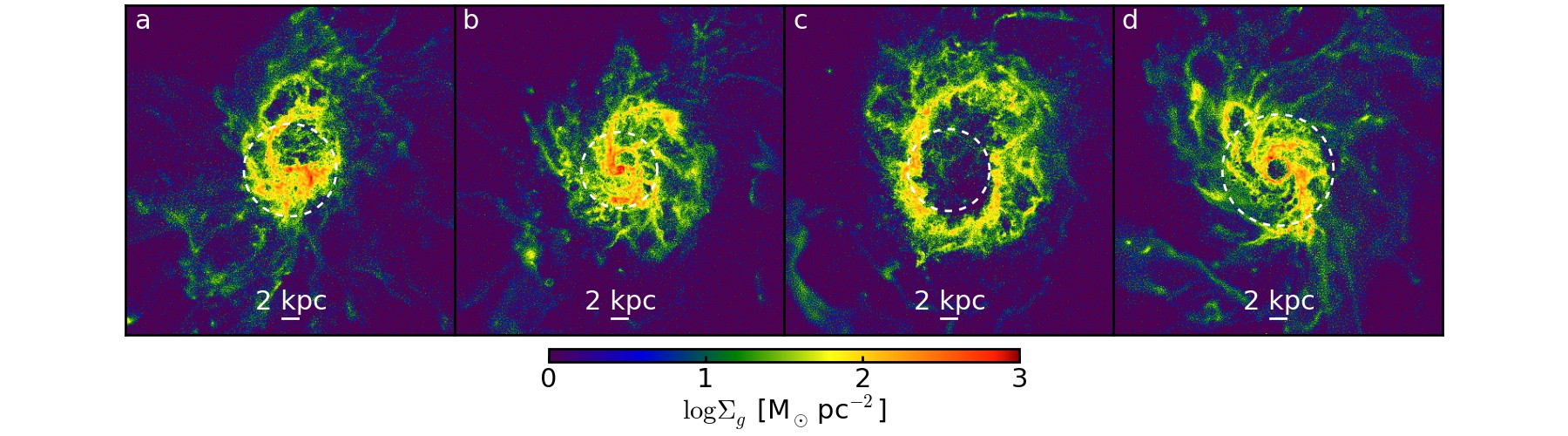} & \\
  \includegraphics[width=\linewidth]{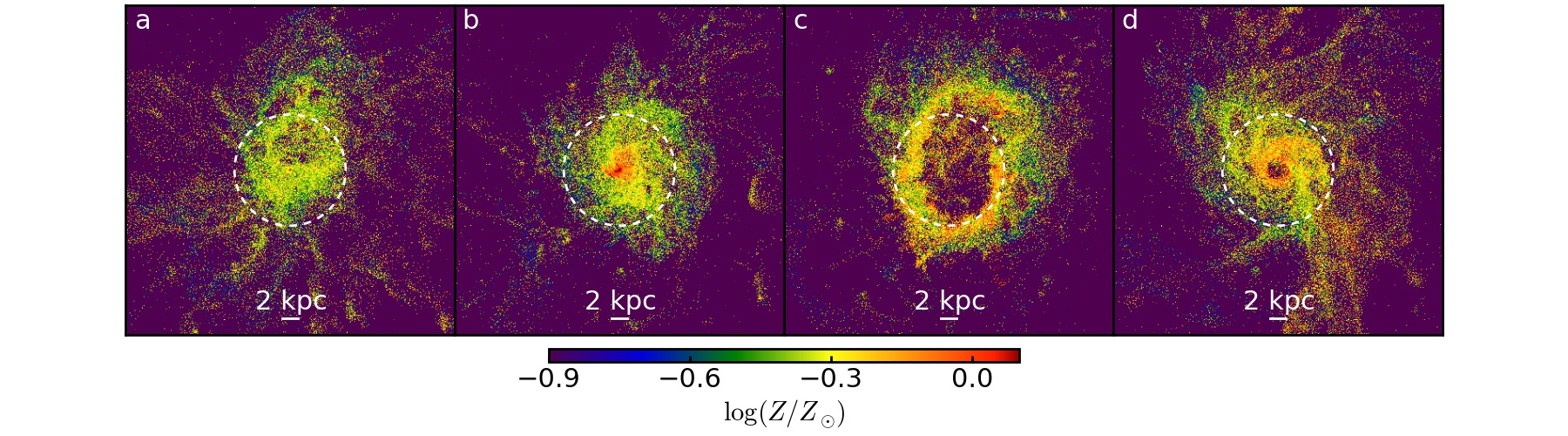} & \\
\end{tabular}
\caption{{\em Top:} Metallicity gradient in the galaxy m12i (measured from 0.25--$1R_{90}$) as a function of cosmic time at redshifts $z=0$--1.1 (black solid). The SFR (red dotted) and gas outflow rate measured at 10\,kpc (blue dashed) are also shown for comparison. {\em Middle:} Gas morphology at the four epochs labeled by the vertical grey dotted lines in the top panel (a--d). {\em Bottom:} Metallicity map at the four epochs. At $z>0.7$, the metallicity gradient shows considerable time fluctuations, associated with starburst episodes. The examples illustrate this process: (a) gas flows in rapidly and forms a disk, (b) a negative metallicity gradient builds up during star formation, (c) strong feedback from starburst drives intense gas outflow and flattens the metallicity gradient, and (d) gas falls back and reforms a disk. The peaks in gas outflow rate match the ``peaks'' in metallicity gradients (where the gradients are flat). This explicitly shows the effect of feedback flattening the metallicity gradient. At $z<0.7$, the disk has `calmed down', and stellar feedback is no longer strong enough to disrupt the gas disk. A negative metallicity gradient then develops rapidly, and does not evolve significantly with time after this.}
\label{fig:m12i}
\end{figure*}

\begin{figure}
\centering
\includegraphics[width=\linewidth]{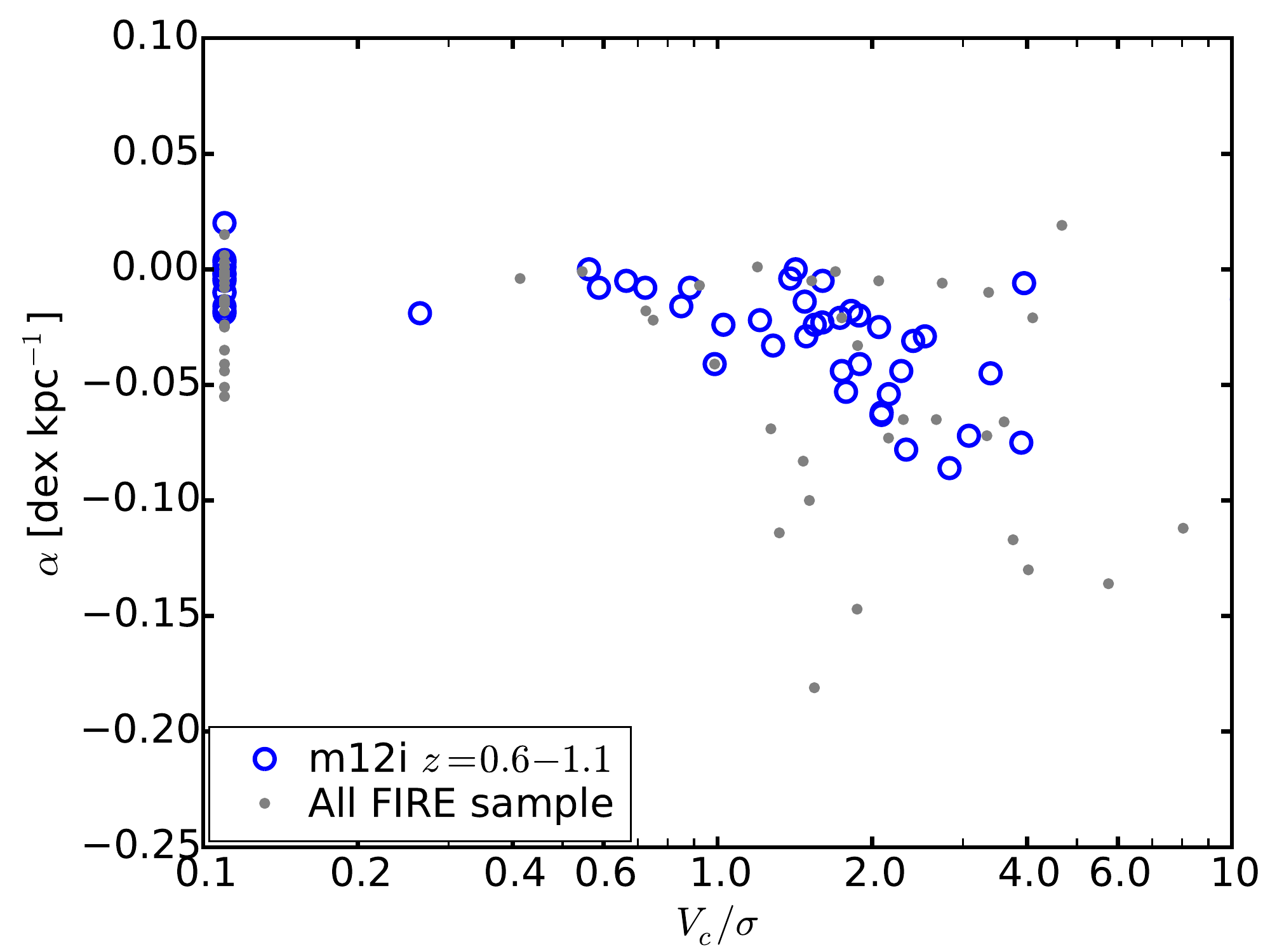}
\caption{Metallicity gradient vs degree of rotational support ($\alpha$--$V_c/\sigma$) for 50 successive snapshots from simulation m12i during $z=0.6$--1.1 (blue circles). The grey points show the entire FIRE sample presented in Figure \ref{fig:alvc}. A single galaxy measured at different time occupies similar parameter space to an ensemble of galaxies -- strong negative metallicity gradients {\em only} appear when there is well-ordered rotation, while the gradient tends to be flat when the galaxy is strongly perturbed. This implies that the observed gradients more closely reflect the instantaneous state of the galaxy than its cosmological growth history.}
\label{fig:alvc_m12i}
\end{figure}

\subsection{Metallicity gradient vs kinematic properties}
\label{sec:zkin}
In the left panel of Fig. \ref{fig:alvc}, we show the relation between gas-phase metallicity gradient (measured over 0.25--$1R_{90}$) and degree of rotational support, $V_c/\sigma$, for the entire simulated sample. Again, galaxies whose $V_c$ cannot be properly determined are plotted at $V_c/\sigma\sim0.1$. In general, our simulated sample can be divided into three populations that occupy three different regions on the $\alpha$--$V_c/\sigma$ diagram: (1) significant negative metallicity gradients {\em only} occur in galaxies with rotationally supported disks ($V_c/\sigma\geq1$), (2) strongly perturbed galaxies, with no evidence of rotation ($V_c/\sigma<1$, including those with undetermined $V_c$), tend to have flat metallicity gradients, and (3) there is also a population that show flat or mildly positive metallicity gradients ($\alpha\sim0$) while being rotationally supported ($V_c/\sigma\geq1$). The existence of  population (3) reflects the observed complex relation between metallicity gradient and galaxy kinematics \citep[e.g.][]{jones.2015:glass.metal.grad,leeth.2016:lense.metal.grad}. We emphasize that our sample only predicts that galaxies with a strong negative metallicity gradient {\em must} be rotationally supported, but {\em not} vice versa. We do {\em not} find any galaxy in our simulated sample that has a significant negative metallicity gradient ($\alpha<-0.05\,\dkpc$) but is strongly perturbed ($V_c/\sigma<1$).

The connection between negative metallicity gradients and rotating disks can be understood from the coevolution of the gas disk and stellar disk \citep[e.g.][]{ho.2015:metal.grad.coevolve}. A simple toy model is useful for illustrative purposes. Start from a pristine gas disk with an exponential surface density profile $\Sg\sim\exp(-R/R_d)$, where $R_d$ is the disk scale length. Stars form in the disk at higher efficiencies in regions with higher surface densities, following the Kennicutt--Schmidt law $\DSs\sim\Sg^{1.4}\sim\exp(-1.4R/R_d)$ \citep{kennicutt.1998:sf.law.review}. If the metals do not mix efficiently between annuli (i.e. the local `closed-box' assumption), the gas-phase metallicity is $Z_g\sim-\ln(1-f_{\ast})$, where $f_{\ast}$ is the mass fraction of stars (note that both $f_{\ast}$ and $Z_g$ are functions of radius). If the gas fraction is not too low, $Z_g\sim f_{\ast}\sim\Ss/\Sigma\sim\DSs t/\Sigma\sim\exp(-0.4R/R_d)$. This naturally gives a negative metallicity gradient $\dd\log Z_g/\dd R=-0.17/R_d\,\dkpc$ (if $R_d$ is in kpc), although the slope can be altered by the exact disk surface density profile\footnote{If the initial gas disk has a power-law surface density profile $\Sg\sim R^{-\beta}$, where $\beta>0$ is the power-law index, following the same argument above, the gas-phase metallicity profile will be $Z_g\sim R^{-0.4\beta}$. A power-law profile might be a better description to our simulations \citep[e.g.][]{hopkins.2009:cusp.ellipticals} and the observed metallicity profiles in early-type galaxies \citep[e.g.][]{reda.2007:resolve.early.type,sanchez.2007:resolve.early.type}. In such case, the slope of metallicity gradients, if defined in $\dd\log Z_g/\dd R$ (in $\dkpc$), also depends on the range where the gradient is measured. This may account for the steep metallicity gradients ($\sim-0.3\,\dkpc$) observed in high-redshift galaxies \citep[e.g.][also see Fig. \ref{fig:alz}]{jones.2013:lense.metal.grad}.}, pre-enrichment in the disk, the strength of radial mixing, etc. Population (2) galaxies are strongly perturbed via violent processes, such as mergers, rapid gas inflows, and strong feedback-driven outflows, which can destroy any pre-existing rotating disk and cause efficient gas re-distribution on galactic scales. Galaxies in region (3) may be in a transition phase, e.g. during a gas infall before a strong negative metallicity gradient builds up at a later time. In Section \ref{sec:feedback}, we will further show that the metallicity gradient and kinematic properties of a galaxy can vary on $\lesssim\Gyr$ time-scales, causing the galaxy to move across the three regions on the $\alpha$--$V_c/\sigma$ relation.

In the right panel of Fig. \ref{fig:alvc}, we show the relation between metallicity gradient and $\Delta V/2\sigma$. Similarly, strong negative metallicity gradients {\em only} appear in galaxies with $\Delta V/2\sigma\geq1$, consistent with the results we find with $V_c/\sigma$. Again, we caution that $\Delta V/2\sigma$ may not be a robust indicator of whether a galaxy is rotationally supported or strongly perturbed (see Fig. \ref{fig:vsigma}). In Fig. \ref{fig:alvc}, we also compare our simulations with observational data from \citet[][Y11]{yuan.2011:metal.gradient.spiral}, \citet[][S12]{swinbank.2012:hizels.metal.grad}, \citet[][J13]{jones.2013:lense.metal.grad}, and \citet[][L16]{leeth.2016:lense.metal.grad}. Note that we follow \citet{leeth.2016:lense.metal.grad} and only adopt the $V_c/\sigma$ for those that can be reliably fitted by a simple disk model ($\chi_{\rm red}^2<20$ in their table 3), while we regard the rest of their sample as non-rotationally supported ($V_c$ undetermined). Our simulations reproduce the observed complexity in the relationship between metallicity gradient and kinematic properties. Remarkably, the simulated sample and the observed sample, although both small in sample size, occupy almost identical parameter space in these relations.

\subsection{Metallicity gradient vs redshift}
\label{sec:zred}
In Fig. \ref{fig:alz}, we plot the metallicity gradients for all simulated galaxies in our sample as a function of redshift, at $z=2$, 1.4, 0.8, and 0. The black points present the metallicity gradients measured from 0.25--$1R_{90}$. We also compare a variety of observations from \citet[][M03]{maciel.2003:mw.metal.grad.evolve}, \citet[][Y11]{yuan.2011:metal.gradient.spiral}, \citet[][S12]{swinbank.2012:hizels.metal.grad}, \citet[][J13]{jones.2013:lense.metal.grad}, \citet[][J15]{jones.2015:glass.metal.grad}, \citet[][L16]{leeth.2016:lense.metal.grad}, and \citet[][W16]{wang.2016:glass.metal.grad}. Our simulations are broadly consistent with the observed diversity of metallicity gradients at redshifts $z=0.5$--2.5. For example, at $z\sim2$, our sample covers metallicity gradients from $\alpha=-0.15$--$0.05\,\dkpc$, in reasonably good agreement with observational data at that epoch. Note that we measure the metallicity gradient from 0.25--$1R_{90}$ by default, whereas there is no universal standard for the radial limits used to define the metallicity gradients in observations. If we instead use the metallicity gradient in the central 0--2\,kpc in our simulations, as shown by the small grey points in Fig. \ref{fig:alz}, we obtain a similar result, but with somewhat larger scatter, with the slope ranging from $-0.3$--$0.1\,\dkpc$. This is in better agreement with the steep slopes and positive metallicity gradients in some of the observational samples \citep[e.g.][]{jones.2013:lense.metal.grad,leeth.2016:lense.metal.grad}. A more rigorous comparison would require matching precisely the galaxy selection function and observational metallicity gradient measurement method of each observed sample, which is beyond the scope of this paper.

We also compare our results with the MUGS simulation (`conservative' feedback) and the MAGICC simulation (`enhanced' feedback) from \citet{gibson.2013:metal.grad.notes}. In the `enhanced' feedback model, gas heated by SNe is kept hot artificially for much longer than the Sedov-Taylor phase to generate efficient outflows \citep{stinson.2013:magicc.method}, in contrast to much simpler `sub-grid' models which effectively suppress bursty star formation. These feedback models also require fine-tuning certain parameters to match the observed galaxy properties. The `conservative' (weak) feedback model in \citet{gibson.2013:metal.grad.notes} always predicts the so-called `inside-out' growth picture. In this scenario, a compact core formed rapidly at the center of the galaxy, building up a steep negative metallicity gradient at high redshift. Then the galaxy gradually grows in size and the metallicity gradient flattens as the galaxy evolves. Their `enhanced' (strong) feedback model, on the other hand, always produces a flat metallicity gradient that shows little evolution with redshift. In contrast, our sample produces more diverse distribution of metallicity gradients in good agreement with observations, including both strong negative gradients and flat/weak positive gradients. 
This confirms that metallicity gradients in cosmological simulations are sensitive to the treatment of feedback. The physics adopted in FIRE explicitly resolves feedback processes on sub-kpc scales which allows galaxies to `switch' between weak and strong outflows based on their local conditions. As a consequence, our simulations produce both strong and weak gradients, even in the same galaxy at slightly different times in its evolution. This leads to a diversity of gradients in good agreement with observations, and in contrast to simpler `sub-grid' feedback models.

\subsection{The effects of feedback: a case study}
\label{sec:feedback}
In this section, we will show how feedback results in the complex relation between galaxy gas-phase metallicity gradients and kinematic properties. To this end, we perform a case study on simulation m12i, which produces a Milky Way-mass disk galaxy by $z=0$. In the top panel of Fig. \ref{fig:m12i}, we show the metallicity gradient (measured from 0.25--$1R_{90}$) as a function of cosmic time at redshifts $z=0$--1.1 (the black solid line). Note that prior to $z=1.1$, this is a clumpy, low-mass galaxy that has chaotic, bursty star formation, with little rotation and flat metallicity gradients \citep{ma.2016:disk.structure.evolve}, so we do not show it here. For comparison, we also show the instantaneous SFR (averaged over 10\,Myr, the red dotted line)\footnote{Note that the SFRs shown here are different from those defined in Section \ref{sec:galaxy} and listed in Appendix \ref{sec:append:gal} (where the SFRs are averaged over 200\,Myr), because we want to emphasize the short-time-scale fluctuations in this section.} and the gas outflow rate at 10\,kpc (the blue dashed line) during the same period. We follow \citet{faucher.2011:dm.halo.assemby} and \citet{muratov.2015:fire.mass.loading} and calculate the gas outflow rate as
\be
\frac{\partial M}{\partial t} = \frac{1}{L} \sum_i m_i \frac{{\bf v}_i \cdot {\bf r}_i}{|{\bf r}_i|},
\ee
where we sum over all gas particles that have radial velocity $v_{\rm r}={\bf v}\cdot{\bf r}/|{\bf r}|>100\,\km\,\s^{-1}$ within the central $L=10\,\kpc$ in the galaxy. 

At $z>0.7$, both the gas outflow rate and SFR show significant time variability. The outflow rates are much higher than the SFRs (high mass loading factors), implying that feedback is very efficient at these times \citep{muratov.2015:fire.mass.loading}\footnote{Note that while the outflow rates in Fig. \ref{fig:m12i} are qualitatively similar to those in \citet{muratov.2015:fire.mass.loading}, they different quantitively because of different radial and velocity range considered.}. At the same time, the metallicity gradient also shows significant fluctuations. Interestingly, the peaks in gas outflow rates coincide with the `peaks' in metallicity gradients (i.e., when the gradient is flat, since a strong gradient has a negative slope). To further illustrate the process, we show example gas images and metallicity maps in the middle and bottom panels in Fig. \ref{fig:m12i}, respectively, at four selected times labeled by (a)--(d), as shown by the grey vertical dotted lines in the top panel of Fig. \ref{fig:m12i}. First, gas flows in rapidly and forms a rotating gas disk (a). Rapid gas infall triggers a starburst in the disk, and a negative metallicity gradient builds up quickly (b, see the argument in Section \ref{sec:zkin}). Next, feedback from the starburst drives strong outflows, which destroy the gas disk and mix the metals on galactic scales, flattening the pre-existing negative metallicity gradient in the disk (c). Finally, gas falls back, reforming a disk, and the next episode starts (d). 

We repeat the analysis in Section \ref{sec:zkin} and measure the degree of rotational support $V_c/\sigma$ for 50 successive snapshots from simulation m12i, from $z=0.6$--1.1, before the metallicity gradient becomes stable. In Fig. \ref{fig:alvc_m12i}, we plot the relation between metallicity gradient and $V_c/\sigma$ for the 50 epochs considered here (blue circles) and compare the results with the entire FIRE sample as shown in Fig. \ref{fig:alvc} (grey points). Remarkably, the time variability of a single galaxy occupies almost identical parameter space as the entire simulated sample in the $\alpha$--$V_c/\sigma$ relation. Again, significant negative metallicity gradients {\em only} appear when there is a well-ordered rotating disk, while the gradients are flat when the galaxy is strongly perturbed and shows little rotation. At the epochs when the galaxy has a flat metallicity gradient but is rotationally supported, it is mostly in the early stage of gas infall before a strong metallicity gradient builds up later (e.g. epoch (a) shown in Fig. \ref{fig:m12i}). These results suggest that a single galaxy can rapidly (in a few 100 Myr) traverse the range of observed metallicity gradients and kinematic properties, indicating that the observed metallicity gradients at high redshifts may be more of an indicator of the {\it instantaneous} ($\lesssim\Gyr$ time-scale) dynamical state of the galaxy, {\it not} the long-term galaxy formation, accretion, or growth history. 

Almost all the simulated galaxies show significant burstiness in SFR and undergo strong bursts of feedback-driven outflows at high redshift ($z\gtrsim0.5$), even for the most massive galaxies at $z\sim2$ \citep{hopkins.2014:fire.galaxy,sparre.2015:fire.sf.burst,muratov.2015:fire.mass.loading}. The central galaxy in simulation m12i calms down after $z\sim0.7$, and there is always a well-ordered, rotationally supported gas disk thereafter \citep{ma.2016:disk.structure.evolve}. Stars form in the disk at a nearly constant rate that is set by the nearly constant gas accretion rate and regulated by stellar feedback. The feedback is no longer sufficient to drive strong gas outflows and destroy the gas disk. A negative metallicity gradient builds up quickly as soon as the disk calms down and stays almost unchanged after this time. A similar transition is also seen in other simulations that produce a galaxy more massive than $\Ms=10^{10}\,\Msun$ by $z=0$, as these galaxies also cannot drive strong gas outflows at late times \citep{muratov.2015:fire.mass.loading}. Such a transition is likely due to a combination of decreasing merger rates at lower redshifts \citep[e.g.][]{hopkins.2010:merger.rates} and decreasing gas fractions in massive galaxies \citep{hayward.hopkins.2015:wind.driving}. Therefore, it is expected that massive galaxies in the local Universe mostly have stable negative metallicity gradients, except for strongly perturbed (e.g. merging) galaxies. 

\section{Discussion and Conclusions}
\label{sec:con}
In this paper, we use 32 high-resolution cosmological zoom-in simulations from the FIRE project to study the gas-phase metallicity gradient in galaxies and its relation with galaxy properties. Our simulated sample includes 32 galaxies at $z=2$, covering a halo mass range $10^{11}$--$10^{13}\,\Msun$ and stellar mass range $10^9$--$10^{11}\,\Msun$. A sub-sample has been run to $z=0$, spanning a halo mass range $10^{11}$--$10^{13}\,\Msun$ and stellar mass range $10^9$--$10^{11}\,\Msun$ at $z=0$. The FIRE simulations include physically motivated models of the multi-phase ISM, star formation, and stellar feedback and have been shown to reproduce a number of observed properties of galaxies for a broad range of stellar mass at redshift $z=0$--6. These simulations explicitly resolve the launching and propagation of galactic winds on sub-kpc scales and can thus capture the effects of stellar feedback on metallicity gradients.
\begin{enumerate}
\item The simulations produce a diverse range of kinematic properties and metallicity gradients, broadly consistent with observations at all redshifts. Our simulated sample includes merging galaxies, starbursts with gas morphologies disturbed by feedback, as well as relatively stable, rotation-dominated disk galaxies. 

\item Strong negative metallicity gradients {\em only} appear in galaxies with a gas disk, as reflected by well-ordered rotation ($V_c/\sigma\geq1$), while strongly perturbed galaxies ($V_c/\sigma<1$) always have flat gradients. In a gas disk, the star formation efficiency is higher toward the center due to increasing gas surface density, so metal enrichment is faster in the central region, leading to a negative metallicity gradient. Strong perturbations driven by rapid gas infall, mergers, or violent outflows, can stir the gas in the ISM, causing metal re-distribution on galactic scales and flattening metallicity gradients. Not {\em all} rotationally supported galaxies have strong negative metallicity gradients.

\item The metallicity gradient and kinematic properties of a high-redshift galaxy can vary on $\lesssim\Gyr$ time-scales, associated with starburst episodes. The time variability of a single galaxy is statistically similar to the overall simulated sample. A negative metallicity gradient can build up quickly as a starburst is triggered in a gas disk formed via gas infall. Strong feedback from the starburst drives intense outflows, which destroy the gas disk and cause metal re-distribution on galactic scales, resulting in flat metallicity gradients. Gas recycles in fountains \citep{angles.2016:fire.baryon.cycle}, and negative gradients may re-establish quickly. This has important consequences for the interpretation of metallicity gradients observed in high-redshift galaxies. They may {\em not} well-correlate with the accretion or growth history of the galaxy on cosmological time-scales, but rather reflect the `instantaneous' state of gas dynamics. 

\item There is weak dependence of metallicity gradient on both stellar mass and sSFR. Low-mass galaxies, and/or galaxies with high sSFR tend to have flat metallicity gradients, owing to efficient feedback in such systems, which keeps them in the `bursty' star formation mode.

\item Because of the important role of stellar feedback, it is essential to resolve feedback from sub-kpc to galactic scales in sufficiently high-resolution simulations, to reproduce the observed diversity of kinematic properties and metallicity gradients in high-redshift galaxies. Our results are in contrast to simulations with simple `sub-grid' feedback models, which tend to predict either `all strong' or `all weak' metallicity gradients.
\end{enumerate}

Our results suggest that the bursty star formation in our simulations can change the kinematic properties and gas-phase metallicity gradients in these galaxies on relatively short time-scales ($\sim10^8$--$10^9$\,yr), which can at least partly explain the diverse kinematics and gradients observed in high-redshift galaxies. One intriguing question we leave open is when and why a galaxy shows such bursty star formation. A detailed answer of this question may require a larger sample of simulations. Nonetheless, the current sample of the FIRE simulations
have suggested that at high redshift ($z>2$), all galaxies show significant burstiness in the SFR, even in the most massive galaxies in the simulated sample \citep{sparre.2015:fire.sf.burst,faucher.2015:fire.neutral.hydrogen,feldmann.2016:massive.fire.long}. At late times, low-mass galaxies ($\Ms<10^{10}\,\Msun$) still have bursty star formation down to $z\sim0$ \citep{wheeler.2015:dwarf.rotation,elbadry.2016:fire.migration}, while more massive galaxies ($\Ms\gtrsim10^{10}\,\Msun$) tend to have a transition from bursty to relatively stable star formation at intermediate redshift \citep[$z\sim0.5$--1,][]{muratov.2015:fire.mass.loading}. \citet{hayward.hopkins.2015:wind.driving} provide an analytic model and argue that such transition happens at a gas fraction threshold of $f_{\rm gas}\sim0.3$, above which the ISM is highly turbulent and star formation is sufficiently violent that feedback can efficiently blow out a large fraction of low-density gas from the disk. At lower gas fractions, turbulence becomes weaker, and feedback is no longer sufficient to drive strong outflows.

In our simulations, stellar metallicity gradients develop coherently with gas-phase metallicity gradients as stars form in the disk (also see the argument in Section \ref{sec:zkin}), but stellar metallicity gradients are much less vulnerable to strong feedback than their gas-phase counterparts, especially in massive galaxies \citep{elbadry.2016:fire.migration}. Stellar migration in the disk can flatten metallicity gradients, but it may only have a weak net effect over a few Gyr time-scale \citep{ma.2016:disk.structure.evolve}. Therefore, we propose that our predictions for the short-time-scale variation of gas-phase metallicity gradients can be tested with stellar metallicity gradients. One would expect that a large fraction of massive high-redshift galaxies have significant negative {\em stellar} metallicity gradients, even if they show a broad range of kinematic properties and gas-phase metallicity gradients. We say massive because the galaxy must have had a gas disk at some point to build up a stellar metallicity gradient, which is not the case in small dwarf galaxies. Negative stellar metallicity gradients have been observed in local galaxies \citep[e.g.][]{sanchez.2014:stellar.grad.califa}, although it is challenging to measure stellar metallicities at higher redshifts. It will be interesting to study stellar metallicity gradients in these simulations in more details in future work.


Nevertheless, our simulations only have a moderate sample size and are limited in statistical power. We show in Section \ref{sec:zkin} that our simulated sample can be divided into three populations based on their kinematic properties and metallicity gradients, but we leave a number of open questions. What fractions of galaxies at a given redshift are rotationally supported and strongly perturbed, respectively? How often are strong perturbations driven by internal feedback vs. external processes? What fraction of rotationally supported galaxies show strong negative gas-phase metallicity gradients? What fraction of galaxies in each population are associated with mergers? These questions are important for understanding high-redshift galaxy populations and worth further investigations, which we hope to explore with larger ensembles of simulations in the future.

\section*{Acknowledgments}
We thank Nicha Leethochawalit, Tucker Jones, Richard Ellis, Xin Wang, and Tommaso Treu for insightful discussions and Nicha Leethochawalit for providing a compilation of observational data. We also acknowledge the anonymous referee for help suggestions on clarifying the manuscript.
The simulations used in this paper were run on XSEDE computational resources (allocations TG-AST120025, TG-AST130039, TG-AST140023, and TG-AST150045) and computational resources provided by the NASA High-End Computing (HEC) Program through the NASA Advanced Supercomputing (NAS) Division at Ames Research Center (proposal SMD-14-5492 and SMD-15-5950).
The analysis was performed on the Caltech compute cluster ``Zwicky'' (NSF MRI award \#PHY-0960291).
Support for PFH was provided by an Alfred P. Sloan Research Fellowship, NASA ATP Grant NNX14AH35G, and NSF Collaborative Research Grant \#1411920 and CAREER grant \#1455342.
 RF was supported in part by NASA through Hubble Fellowship grant HF2-51304.001-A awarded by the Space Telescope Science Institute, which is operated by the Association of Universities for Research in Astronomy, Inc., for NASA, under contract NAS 5-26555, in part by the Theoretical Astrophysics Center at UC Berkeley, and by NASA ATP grant 12-ATP-120183.
CAFG was supported by NSF through grants AST-1412836 and AST-1517491, by NASA through grant NNX15AB22G, and by STScI through grant HST-AR-14293.001-A.
DK was supported by NSF grant AST-1412153 and Cottrell Scholar Award from the Research Corporation for Science Advancement. 

\bibliography{../../zlib}

\appendix

\section{Galaxy properties}
\label{sec:append:gal}

\begin{table*}
\caption{Galaxy properties, kinematics, and metallicity gradients of the simulated sample.}
\centering
\begin{center}
\begin{tabular}{ccccccccc}
\hline
Name & $z$ & $M_{\ast}$ & SFR & $R_{90}$ & $V_c$ & $\Delta V/2$ & $\sigma$ & $\alpha$ \\
 & & ($\Msun$) & ($\Msun\,\yr^{-1}$) & (kpc) & ($\km\,\s^{-1}$) & ($\km\,\s^{-1}$) & ($\km\,\s^{-1}$) & ($\dex\,\kpc^{-1}$) \\ 
\hline
m11 & 2.0 & 1.9e8 & -- & -- & -- & -- & -- & -- \\
m12v & 2.0 & 1.0e9 & 0.05 & 4.37 & -- & 76.7 & 77.5 & $-0.055\pm0.007$  \\ 
m12q & 2.0 & 3.0e9 & 0.70 & 2.51 & 85.4 & 55.8 & 37.2 & $-0.065\pm0.007$  \\ 
m12i & 2.0 & 5.3e8 & 0.52 & 4.57 & 44.2 & 30.6 & 16.2 & $-0.006\pm0.005$  \\ 
m13 & 2.0 & 2.0e10 & 2.2 & 3.07 & 157.8 & 168.8 & 84.5 & $-0.147\pm0.008$ \\
m11h383 & 2.0 & 3.5e8 & 0.04 & 2.8 & -- & 8.3 & 12.0 & $-0.051\pm0.007$ \\
m11.4a & 2.0 & 4.6e8 & 0.17 & 3.4 & -- & 16.6 & 29.6 & $-0.044\pm0.006$ \\
m11.9a & 2.0 & 6.3e8 & 0.16 & 4.1 & -- & 21.3 & 20.9 & $-0.015\pm0.005$ \\
MFz0\_A2 & 2.0 & 1.1e11 & 55.6 & 4.6 & 482.4 & 365.6 & 120.0 & $-0.130\pm0.008$ \\
z2h350 & 2.0 & 6.4e9 & 9.4 & 3.44 & -- & -- & -- & -- \\
z2h400 & 2.0 & 5.8e9 & 4.3 & 4.89 & 67.9 & 106.8 & 93.5 & $-0.018\pm0.002$ \\
z2h450 & 2.0 & 6.7e9 & 0.35 & 9.42 & -- & 114.1 & 103.2 & $-0.004\pm0.002$ \\
z2h506 & 2.0 & 8.1e9 & 5.4 & 7.85 & 116.5 & 39.4 & 28.4 & $-0.021\pm0.003$ \\
z2h550 & 2.0 & 9.6e8 & 0.85 & 3.77 & -- & 25.6 & 25.5 & $-0.018\pm0.004$ \\
z2h600 & 2.0 & 1.1e10 & 6.2 & 7.67 & -- & 59.6 & 35.8 & $-0.001\pm0.002$ \\
z2h650 & 2.0 & 5.2e9 & 4.4 & 6.46 & -- & 31.1 & 12.5 & $-0.013\pm0.003$ \\
z2h830 & 2.0 & 5.1e9 & 2.6 & 5.32 & -- & 3.0 & 7.8 & $-0.029\pm0.011$ \\
A1:0 & 2.0 & 2.3e10 & 9.9 & 2.90 & 72.4 & 85.5 & 132.7 & $-0.001\pm0.013$ \\
A2:0 & 2.0 & 3.1e10 & 14.0 & 6.88 & 171.7 & 167.7 & 91.7 & $-0.033\pm0.003$ \\
A3:0 & 2.0 & 1.1e10 & 6.9 & 2.98 & 148.8 & 128.4 & 117.2 & $-0.069\pm0.008$ \\
A4:0 & 2.0 & 1.2e10 & 2.8 & 2.43 & 167.6 & 158.2 & 111.1 & $-0.100\pm0.015$ \\
A5:0 & 2.0 & 1.6e10 & 28.5 & 6.67 & -- & 115.5 & 66.4 & $0.007\pm0.004$ \\
A6:0 & 2.0 & 2.3e10 & 1.9 & 5.83 & 50.5 & 159.8 & 122.3 & $-0.004\pm0.003$ \\
A7:0 & 2.0 & 2.0e10 & 12.9 & 8.02 & -- & 67.6 & 103.0 & $-0.003\pm0.001$ \\
A8:0 & 2.0 & 1.1e10 & 10.1 & 7.46 & -- & 67.0 & 118.0 & $-0.006\pm0.001$ \\
A9:0 & 2.0 & 8.3e9 & 3.9 & 3.80 & 23.9 & 20.0 & 25.9 & $-0.007\pm0.007$ \\
A10:0 & 2.0 & 2.3e10 & 22.5 & 6.62 & 76.1 & 64.3 & 49.9 & $-0.003\pm0.004$ \\
B1:0 & 2.0 & 6.8e10 & 38.5 & 6.50 & -- & 253.7 & 235.4 & $-0.035\pm0.002$ \\
B2:0 & 2.0 & 6.0e10 & 6.2 & 7.30 & -- & 280.3 & 248.0 & $-0.025\pm0.002$ \\
B3:0 & 2.0 & 5.0e10 & 40.5 & 8.26 & 428.0 & 323.9 & 199.1 & $-0.073\pm0.003$ \\
B4:0 & 2.0 & 2.5e10 & 45.3 & 8.00 & -- & -- & -- & -- \\
B5:0 & 2.0 & 4.1e10 & 14.2 & 5.39 & 202.5 & 143.0 & 138.0 & $-0.083\pm0.002$ \\
\hline
\multicolumn{9}{p{0.7\linewidth}}{Galaxy properties studied in this paper (units are physical):} \\
\multicolumn{9}{p{0.7\linewidth}}{(1) Name: Simulation designation.} \\
\multicolumn{9}{p{0.7\linewidth}}{(2) $z$: redshift where the properties here are measured.} \\
\multicolumn{9}{p{0.7\linewidth}}{(3) $\Ms$: Stellar mass within the central 10\,kpc of the galaxy at the given redshift.} \\
\multicolumn{9}{p{0.7\linewidth}}{(4) SFR: Star formation rate within the central 10\,kpc of the galaxy (averaged over 200\,Myr).} \\ 
\multicolumn{9}{p{0.7\linewidth}}{(5) $R_{90}$: Defined in Section \ref{sec:galaxy}, as the radius that encloses 90\% of the stars younger than 200\,Myr within 10\,kpc.}\\
\multicolumn{9}{p{0.7\linewidth}}{(6) $V_c$: Rotation velocity given by the arctan fit from Equation (\ref{eqn:arctan}) to the gas velocity curve (see Section \ref{sec:kin}).} \\ 
\multicolumn{9}{p{0.7\linewidth}}{(7) $\Delta V$: Peak-to-peak velocity difference in the gas velocity curve (see Section \ref{sec:kin}).} \\
\multicolumn{9}{p{0.7\linewidth}}{(8) $\sigma$: Maximum velocity dispersion of gas (see Section \ref{sec:kin}).} \\
\multicolumn{9}{p{0.7\linewidth}}{(9) $\alpha$: Gas-phase metallicity gradient measured over 0.25--$1R_{90}$ from Equation (\ref{eqn:zgrad}).} \\
\multicolumn{9}{p{0.7\linewidth}}{Note: If a galaxy is temporarily quenched and near gas depletion in the central 10\,kpc, its gas kinematic properties ($V_c$, $\Delta V/2$, and $\sigma$) and gas-phase metallicity gradient ($\alpha$) cannot be properly measured. If a galaxy has been quenched for more than 200\,Myr, its SFR and $R_{90}$ are also not defined.} \\
\end{tabular}
\end{center}
\label{tbl:param}
\end{table*}%

\begin{table*}
\caption{Galaxy properties, kinematics, and metallicity gradients of the simulated sample. --- Continued.}
\centering
\begin{center}
\begin{tabular}{ccccccccc}
\hline
Name & Redshift & $M_{\ast}$ & SFR & $R_{90}$ & $V_c$ & $\Delta V/2$ & $\sigma$ & $\alpha$ \\
 & & ($\Msun$) & ($\Msun\,\yr^{-1}$) & (kpc) & ($\km\,\s^{-1}$) & ($\km\,\s^{-1}$) & ($\km\,\s^{-1}$) & ($\dex\,\kpc^{-1}$) \\ 
\hline
m11 & 1.4 & 3.4e8 & 0.12 & 7.1 & 30.6 & 16.0 & 6.6 & $0.019\pm0.003$ \\
m12v & 1.4 & 7.9e9 & 6.0 & 7.36 & 83.2 & 71.8 & 49.2 & $-0.001\pm0.001$  \\ 
m12q & 1.4 & 8.4e9 & 5.1 & 3.68 & 119.0 & 83.5 & 68.2 & $-0.021\pm0.004$  \\ 
m12i & 1.4 & 6.0e9 & 1.7 & 4.44 & 5.7 & 52.2 & 67.9 & $0.003\pm0.004$  \\ 
m13 & 1.4 & 5.8e10 & 14.9 & 5.39 & 344.3 & 251.0 & 59.8 & $-0.136\pm0.006$ \\
m11h383 & 1.4 & 4.7e8 & 0.14 & 1.9 & -- & 61.8 & 70.4 & $-0.041\pm0.011$ \\
m11.4a & 1.4 & 5.3e8 & 0.35 & 2.5 & 17.3 & 15.9 & 23.0 & $-0.022\pm0.008$ \\
m11.9a & 1.4 & 1.9e9 & 1.4 & 8.1 & -- & 70.4 & 27.3 & $0.001\pm0.001$ \\
MFz0\_A2 & 1.4 & 1.3e11 & 11.5 & 4.3 & 537.6 & 392.8 & 123.7 & $-0.114\pm0.008$ \\
m11 & 0.8 & 8.4e8 & 0.05 & 4.8 & -- & 14.3 & 22.7 & $0.003\pm0.005$ \\
m12v & 0.8 & 9.1e9 & 0.004 & 4.13 & 133.8 & 108.8 & 39.8 & $-0.010\pm0.004$  \\ 
m12q & 0.8 & 1.1e10 & 0.001 & 4.56 & -- & -- & -- & -- \\ 
m12i & 0.8 & 1.3e10 & 10.1 & 4.12 & 76.6 & 88.0 & 77.6 & $-0.041\pm0.004$  \\ 
m13 & 0.8 & 6.4e10 & 2.4 & 7.04 & 435.7 & 251.0 & 54.2 & $-0.112\pm0.013$ \\
m11h383 & 0.8 & 1.2e9 & 0.01 & 2.6 & -- & 27.3 & 10.9 & $0.015\pm0.010$ \\
m11.4a & 0.8 & 1.8e9 & 0.07 & 6.4 & 62.3 & 53.1 & 30.3 & $-0.005\pm0.002$ \\
m11.9a & 0.8 & 3.0e9 & 0.4 & 4.4 & -- & 41.6 & 26.8 & $-0.015\pm0.004$ \\
MFz0\_A2 & 0.8 & 1.4e11 & 4.1 & 5.3 & 411.9 & 352.4 & 109.7 & $-0.117\pm0.004$ \\
m11 & 0 & 1.9e9 & 0.4 & 6.4 & -- & 24.1 & 15.3 & $-0.002\pm0.002$ \\
m12v & 0 & 2.2e10 & 0.65 & 4.28 & 141.7 & 161.0 & 91.9 & $-0.181\pm0.008$  \\ 
m12q & 0 & 1.5e10 & 0.50 & 7.71 & 141.8 & 129.8 & 53.3 & $-0.065\pm0.005$  \\ 
m12i & 0 & 4.7e10 & 5.4 & 8.35 & 215.2 & 180.4 & 64.4 & $-0.072\pm0.002$ \\ 
m13 & 0 & 8.2e10 & 1.2 & 3.81 & 324.3 & 242.3 & 89.9 & $-0.066\pm0.005$ \\
m11h383 & 0 & 2.9e9 & 0.3 & 5.0 & -- & 37.2 & 27.0 & $-0.008\pm0.002$ \\
m11.4a & 0 & 4.1e9 & 0.1 & 8.8 & 55.2 & 51.2 & 46.1 & $0.001\pm0.001$ \\
m11.9a & 0 & 1.4e10 & 1.8 & 8.4 & -- & 59.6 & 39.2 & $-0.001\pm0.002$ \\
MFz0\_A2 & 0 & 1.5e11 & -- & -- & -- & -- & -- & -- \\
\hline
\end{tabular}
\end{center}
\label{tbl:paramcon}
\end{table*}%

In this section, we list the galaxy properties (stellar mass, star formation rate, and $R_{90}$, Section \ref{sec:galaxy}), kinematic properties ($V_c$, $\Delta V/2$, and $\sigma$, Section \ref{sec:kin}), and gas-phase metallicity gradient measured in 0.25--$1R_{90}$ (Section \ref{sec:grad}), for the entire simulated sample studied in this paper (Figs. \ref{fig:alms}--\ref{fig:alz}). A machine-readable version of this table is available at \href{http://www.tapir.caltech.edu/~xchma/data/metal_grad.txt}{http://www.tapir.caltech.edu/{\textasciitilde}xchma/data/metal\_grad.txt}.

\label{lastpage}

\end{document}